\documentclass[twocolumn,tighten]{aastex631}
\usepackage{amsmath,amstext}
\usepackage[T1]{fontenc}
\usepackage{ae,aecompl}
\usepackage[utf8]{inputenc}
\usepackage[figure,figure]{hypcap}
\usepackage{enumitem}
\usepackage{bm}
\usepackage{xcolor}
\reportnum{DES-2023-0789}
\reportnum{FERMILAB-PUB-23-743}

\DeclareRobustCommand{\VAN}[3]{#2}
\let\VANthebibliography\thebibliography
\def\thebibliography{\DeclareRobustCommand{\VAN}[3]{##3}\VANthebibliography}

\newcommand{\sbfu}{\rm{mag\,arcsec}^{-2}}
\newcommand{\gi}{g-i}
\newcommand{\zm}{z0MGS}
\newcommand{\mug}{\mu_{\rm eff,g}}
\defcitealias{2021ApJS..252...18T}{T21}

\shorttitle{Quenching of LSBGs}
\shortauthors{Bhattacharyya et al.}

\begin{document}

\title{Environmental Quenching of Low Surface Brightness Galaxies near Milky Way mass Hosts}

\correspondingauthor{Joy Bhattacharyya}
\email{bhattacharyya.37@osu.edu}

%\label{firstpage}

\collaboration{80}{Dark Energy Survey Collaboration:}

\author[0000-0001-6442-5786]{J.~Bhattacharyya$^{1,2}$}
\author{A.H.G.~Peter$^{1,3,2,4}$}
\author{P.~Martini$^{1,3,2}$}
\author{B.~Mutlu-Pakdil$^{5,6,7}$}
\author{A.~Drlica-Wagner$^{6,7,8}$}
\author{A.B.~Pace$^{9}$}
\author{L.E.~Strigari$^{10}$}
\author{Y.-T.~Cheng$^{11}$}
\author{D.~Roberts$^{2}$}
\author{D.~Tanoglidis$^{7}$}
\author{M.~Aguena$^{12}$}
\author{O.~Alves$^{13}$}
\author{F.~Andrade-Oliveira$^{13}$}
\author{D.~Bacon$^{14}$}
\author{D.~Brooks$^{15}$}
\author{A.~Carnero~Rosell$^{16,12,17}$}
\author{J.~Carretero$^{18}$}
\author{L.~N.~da Costa$^{12}$}
\author{M.~E.~S.~Pereira$^{19}$}
\author{T.~M.~Davis$^{20}$}
\author{S.~Desai$^{21}$}
\author{P.~Doel$^{15}$}
\author{I.~Ferrero$^{22}$}
\author{J.~Frieman$^{23,7}$}
\author{J.~Garc\'ia-Bellido$^{24}$}
\author{G.~Giannini$^{18,7}$}
\author{D.~Gruen$^{25}$}
\author{R.~A.~Gruendl$^{26,27}$}
\author{S.~R.~Hinton$^{20}$}
\author{D.~L.~Hollowood$^{28}$}
\author{K.~Honscheid$^{2,3}$}
\author{D.~J.~James$^{29}$}
\author{K.~Kuehn$^{30,31}$}
\author{J.~L.~Marshall$^{32}$}
\author{J. Mena-Fern{\'a}ndez$^{33}$}
\author{R.~Miquel$^{34,18}$}
\author{A.~Palmese$^{35}$}
\author{A.~Pieres$^{12,36}$}
\author{A.~A.~Plazas~Malag\'on$^{37,38}$}
\author{E.~Sanchez$^{39}$}
\author{B.~Santiago$^{40,12}$}
\author{M.~Schubnell$^{13}$}
\author{I.~Sevilla-Noarbe$^{39}$}
\author{M.~Smith$^{41}$}
\author{E.~Suchyta$^{42}$}
\author{M.~E.~C.~Swanson$^{26}$}
\author{G.~Tarle$^{13}$}
\author{M.~Vincenzi$^{14,41}$}
\author{A.~R.~Walker$^{43}$}
\author{N.~Weaverdyck$^{13,44}$}
\author{and P.~Wiseman$^{41}$}

\collaboration{80}{\footnotesize \textit{(Affiliations are shown at the end of this paper.)}}

\begin{abstract}

%Low Surface Brightness Galaxies (LSBGs) are excellent tracers to probe the structures around massive galaxies. Here we study LSBGs extensively across.. with $\mug >24.2\,\sbfu$ drawn from the Dark Energy Survey Y3 catalog that are located near Milky Way mass and lower mass host galaxies in the local universe. These hosts are selected from the z = 0 Multiwavelength Galaxy Synthesis (z0MGS) sample and are spread across a diverse range of environments ranging from the Fornax-Eridanus cluster to isolated field galaxies.

Low Surface Brightness Galaxies (LSBGs) are excellent probes of quenching and other environmental processes near massive galaxies. We study an extensive sample of LSBGs near massive hosts in the local universe that are distributed across a diverse range of environments. The LSBGs with surface-brightness $\mug >24.2\,\sbfu$ are drawn from the Dark Energy Survey Year 3 catalog while the hosts with masses $9.0<{\rm log}(\mathcal{M}_{\star}/M_{\odot})<11.0$ comparable to the Milky Way and the Large Magellanic Cloud are selected from the z0MGS sample. We study the projected radial density profiles of LSBGs as a function of their color and surface brightness around hosts in both the rich Fornax-Eridanus cluster environment and the low-density field. We detect an overdensity with respect to the background density, out to 2.5 times the virial radius for both hosts in the cluster environment and the isolated field galaxies. When the LSBG sample is split by $\gi$ color or surface brightness $\mug$, we find the LSBGs closer to their hosts are significantly redder and brighter, like their high surface-brightness counterparts. The LSBGs form a clear ``red sequence” in both the cluster and isolated environments that is visible beyond the virial radius of the hosts. This suggests a pre-processing of infalling LSBGs and a quenched backsplash population around both host samples. However, the relative prominence of the ``blue cloud" feature implies that pre-processing is ongoing near the isolated hosts compared to the cluster hosts. 

\end{abstract}

\section{Introduction} \label{sec:intro}

Low-surface brightness galaxies (LSBGs) are fainter in terms of their surface brightness relative to the brightness of the night-sky \citep[for a review see][]{1997ARA&A..35..267I}. This fundamental characteristic \citep{1995AJ....110..573M} rendered it difficult to completely survey their population at the surface brightness limit of older surveys \citep{1997AJ....114..635D}. This means that there remains a vast potential for discovery, characterization, and understanding of how they fit in existing models of galaxy formation and evolution. Starting from pioneering works (e.g. \citet{1984AJ.....89..919S,1989AJ.....98..367F,1994AJ....107..530M,2001ApJ...552L..23D,2003A&A...412...45P}), samples of LSBGs are now larger and extend deeper with the technological leaps that has been enabled by digital surveys. Expansive catalogs of such objects  \citep{2012ApJS..200....4F,2015ApJ...813L..15M,2018ApJ...857..104G,2021ApJS..252...18T} have been observed using the CFHT-MegaCam, Dark Energy Survey \citep[DES,][]{2005astro.ph.10346T} and the Hyper-Suprime Cam Strategic Survey Program \citep[HSC-SSP,][]{2018PASJ...70S...4A}.

%\textbf{We want to learn the processing/evolution/quenching of LSBGs$- $ specific question (unknowns), foreshadowing}
These LSBGs constitute a heterogenous population with diversity in both effective radii and luminosity (see Figure 12 in \citet{2018ApJ...857..104G}). For example, LSBGs which have sizes similar to classical dwarf spheroidal (dSph) galaxies contain the subset of ultra-faint dwarfs (UFDs) \citep{2019ARA&A..57..375S} that have extremely low luminosity. LSBGs also include outliers in surface brightness of more extended galaxies, like the ultra-diffuse dwarf galaxies (UDGs) \citep{2015ApJ...798L..45V,2015ApJ...807L...2K,2017MNRAS.468.4039R,2017ApJ...834...16M}. Furthermore, all of these objects have been found across a range of environments ranging from those in the Local Group \citep[e.g.][]{2022arXiv220311788C,2022MNRAS.515L..72C} to those in clusters \citep[e.g.][]{2015ApJ...798L..45V,2015ApJ...809L..21M,2015ApJ...807L...2K} and fields \citep[e.g.][]{2022ApJ...935L..17S}. The ubiquity of these galaxies lead us to question how they might have evolved, if their environments played a role in shaping them and if their evolution differ from that of their brighter counterparts.

Satellite galaxies are shaped by virtue of their close proximity to massive host galaxies that can alter the satellites' gas, stellar and dark matter components \citep[e.g.][]{2010ApJ...721..193P,2014MNRAS.444.2938H}. This causes the morphologies of satellites to differ from other low mass field galaxies, e.g., the bimodal color distribution of bright satellites \citep{2001AJ....122.1861S,2004ApJ...600..681B,2004ApJ...615L.101B,2006MNRAS.373..469B,2009ARA&A..47..159B}. The optical band primarily traces the distribution of stars in a galaxy, which in turn is responsible for its morphology. A crucial way in which the host can alter the stellar component of the satellite is quenching, wherein the galaxy's reservoir of cold gas is removed on account of dynamical interaction with the host.

The low mass, low surface brightness satellites of massive central hosts have also been particularly well surveyed around the Milky Way (MW) and Andromeda \citep[e.g.][]{2012AJ....144....4M}, Local Volume \citep[ELVES,][]{2021ApJ...922..267C} and MW analogs \citep[SAGA,][] {2017ApJ...847....4G}. Such studies have resulted in a good understanding of how the process of quenching operates for dwarf satellites inside cluster and group environments \citep{2013MNRAS.432..336W,2014MNRAS.439.2687W} as well as MW sized halos \citep{2022MNRAS.514.5276S,2022MNRAS.511.1544F,2022arXiv220813805P}. Among the various processes at play it is believed that ram-pressure stripping \citep{1972ApJ...176....1G} by the dense gas of the host is essential in quenching satellites.

%At this point a question arises about what would constitute the spatial extent of the host galaxy in terms of classifying a dwarf as a satellite or a field galaxy. 

At this point a question arises about the extent up to which a massive host can influence the evolution of a dwarf , i.e., classifying one as a satellite or field galaxy. Based on the relative abundances of quenched and star-forming galaxies, it is reasonable to call the dwarfs beyond 1.5 Mpc from massive hosts as the field population  \citep{2012ApJ...757...85G}. Whereas satellites, bound to the host halo are located within the virial radius, which for a MW analogous host is $\sim 200$ kpc. Between these two scales of hostcentric distances, quenching of galaxies that would classically be classified as neither satellite nor field galaxies, deserve more attention \citep{2009ApJ...697..247W,2014MNRAS.439.2687W}. Quenched galaxies which are situated outside the virial radius have been found near cluster mass hosts \citep{2000ApJ...540..113B}, and for less massive systems as well \citep{2018MNRAS.478..548S}. Cosmological simulations show that their quenching took place during their pericentric passages and they currently are outside the virial radius by virtue of their eccentric orbits \citep{2005MNRAS.356.1327G,2009ApJ...692..931L,2009ApJ...697..247W,2018MNRAS.477.4491F,2021NatAs.tmp..160B,2021ApJ...909..112D}. 
 
Galaxies beyond the virial radius of more massive hosts can be split into populations$- $ the backsplash galaxies that previously made a pericenter passage inside the virial radius and the infalling galaxies which are on their first infall \citep{2021MNRAS.501.5948B}. The latter can be subject to pre-processing \citep{2017MNRAS.467.3268R} when accreted as a low mass group among various other modes of quenching driven by the environment of the host and the two galaxy types resemble each other \citep{2011MNRAS.412..529K}. Henceforth in this work we will use the term ``associated galaxies" \citet{2009ApJ...692..931L} to collectively refer to these two populations that interact with the massive hosts yet would not be classified as satellites. 

Contemporary studies of associated galaxies have been biased in favor of those which are bright and exist in high density environment like clusters. However, a picture is emerging wherein LSBGs potentially represent a large proportion of the associated galaxies \citep{2021ApJ...906...96A,2021A&A...656A..44R,2022arXiv221011070K} with signatures of quenching around those near massive galaxies observed as well \citep{2021ApJS..252...18T,2021MNRAS.500.2049P,2022arXiv220502193Z,2022arXiv220411883E}. Since we understand that LSBGs may have diverse modes of formation and subsequent evolution \citep{1997AJ....114..635D,2019MNRAS.485..796M,2021ApJ...920...72K}, it is imperative that we study quenching of LSBGs beyond the virial region of massive galaxies for it can inform us how they fit in the scheme of galaxy evolution. A better understanding can also shed light on the dark matter \citep{2022arXiv220710638A} and baryonic physics \citep{2017MNRAS.466L...1D}.

This paper particularly focuses on how environmental quenching \citep{2018MNRAS.477.4491F} affects LSBGs. We use the Y3 DES catalog of 23790 LSBGs \citep{2021ApJS..252...18T} and investigate how they cluster with respect to 2034 host galaxies in the local universe with masses in the range of Large Magellanic Cloud (LMC) to the Milky Way that are drawn from the z = 0 Multiwavelength Galaxy Synthesis (\zm) catalog of \citet{2019ApJS..244...24L}. We associate LSBGs with hosts through their projected separations which enables us to identify populations of satellite and associated LSBGs. Studying their photometric properties after statistically subtracting any background contribution, we are able to characterize these LSBGs. In order to explore the environment dependency of quenching, we divide the hosts into those that live in the Fornax-Eridanus cluster region and those that are isolated in the field. Using radial density profiles and color-surface brightness distributions, we identify the signatures of quenching beyond the virial radius of the hosts and we contrast them between these two environments. 

%This is not only the first study of backsplash LSBGs but also the first systematic study of LSBG quenching in less dense environments. 

In \S \ref{sec:lsbg_sample} we describe the LSBG sample and its properties. In \S \ref{sec:zm_samp} we describe how we select the host sample of \zm{} galaxies, classify them according to their environments and connect them to the LSBG sample. In \S \ref{sec:analysis} we detail our analysis and results. Ultimately in \S \ref{sec:discuss} we put these findings in the context of current understanding of galaxy evolution and make some suggestions for further work.

\section{Low-surface Brightness Galaxies and their Properties} \label{sec:lsbg_sample}

\subsection{Sample} 

%When considering which target sample of LSBGs to use, those generated from contemporary surveys including HSC-SSP \citep{2018ApJ...857..104G} and FDS \citep{2017A&A...608A.142V} were available but for the purpose of this work 
For this work, we use the Shadows in the Dark sample \citep{2021ApJS..252...18T} of LSBGs that was derived from the DES Y3 data. This constitutes the most extensive catalog both in number and area, with 23790 objects spread over 5000 deg$^2$ of the DES footprint. Since we intend to cross-correlate with host galaxies in the local universe, a large area of the survey footprint is preferred. The fact that the LSB nature of these objects limit the depth of the sample, the large footprint of DES is appropriate in maximizing this volume.

%To prepare this sample, a raw catalog was assembled by applying \texttt{SourceExtractor} on DES Y3 photometric data \citep{2021ApJS..254...24S}. 

In Shadows in the Dark, LSBGs were selected using cuts in the g-band half-light radius $2.5''< r_{1/2,g}< 20''$ and mean surface-brightness $24.2 <\mu_{\rm eff,g}< 28.8\,\sbfu$. In the size-luminosity space, as seen in Figure 15 of \citet{2021ApJS..252...18T}, this sample overlaps with the dwarf galaxies in the Next Generation Fornax Survey (NGFS) catalog \citep{2015ApJ...813L..15M}. This overlap shows that our LSBG sample consists of the type of dwarf galaxies we are interested in$- $ namely low redshift LSBGs that are satellites and associated galaxies around more massive galaxies.

%This was followed by a linear support vector machines (SVM) based classifier algorithm that filtered out potential contaminants like tidal debris and galactic cirrus. Visual inspection and 

The completeness of the Shadows in the Dark sample was established in \citet{2021ApJ...920...72K} by comparing the surface brightness distribution with the HSC sample \citet{2018ApJ...857..104G}, which is deeper and has well defined completeness. Beyond $\mug > 25.75\,\sbfu$, the completeness of the DES sample falls below that of the HSC sample (see Figure 2 of \citet{2021ApJ...920...72K}), which establishes this as the limit of its 80\% completeness. \citet{2021ApJS..252...18T} matched this sample to the Fornax Deep Survey (FDS) catalog \citep{2017A&A...608A.142V} of dwarfs with the same cuts on surface brightness and size as the DES sample applied and determined the completeness in the Fornax region $\sim 66 \%$. They also found that the cuts in the LSBG angular sizes used to define the sample is biased towards the selection of more distant LSBGs with large physical galaxies. The same choice of size cuts also imply that the catalog is incomplete in terms of the extended UDGs and most of the member LSBGs have sizes comparable to classical dwarfs.

\subsection{Photometric properties}

The photometric parameters for each LSBG is derived in \citet{2021ApJS..252...18T} from Sérsic fits made using \texttt{galfitm} \citep{2002AJ....124..266P,2013MNRAS.430..330H}. The effective radius $R_{\rm eff}$ of the LSBGs in this catalog is the semi major exis $a$ of the isophotal ellipse that contains half of the flux from the Sérsic model. However, the mean surface brightness $\mu_{\rm eff}$ is defined as the average light inside the circularized effective radius that is the geometric mean of both axes of the ellipse, $\sqrt{a \ b}$. We use the $\mu_{\rm eff}$ parameter rather than the central surface brightness $\mu_0$ because the latter is often biased by the presence of nuclear star clusters \citep{2020ApJ...902...45S,2022ApJ...927...44C} 

\begin{figure}
\centering
\includegraphics[scale=0.4,clip]{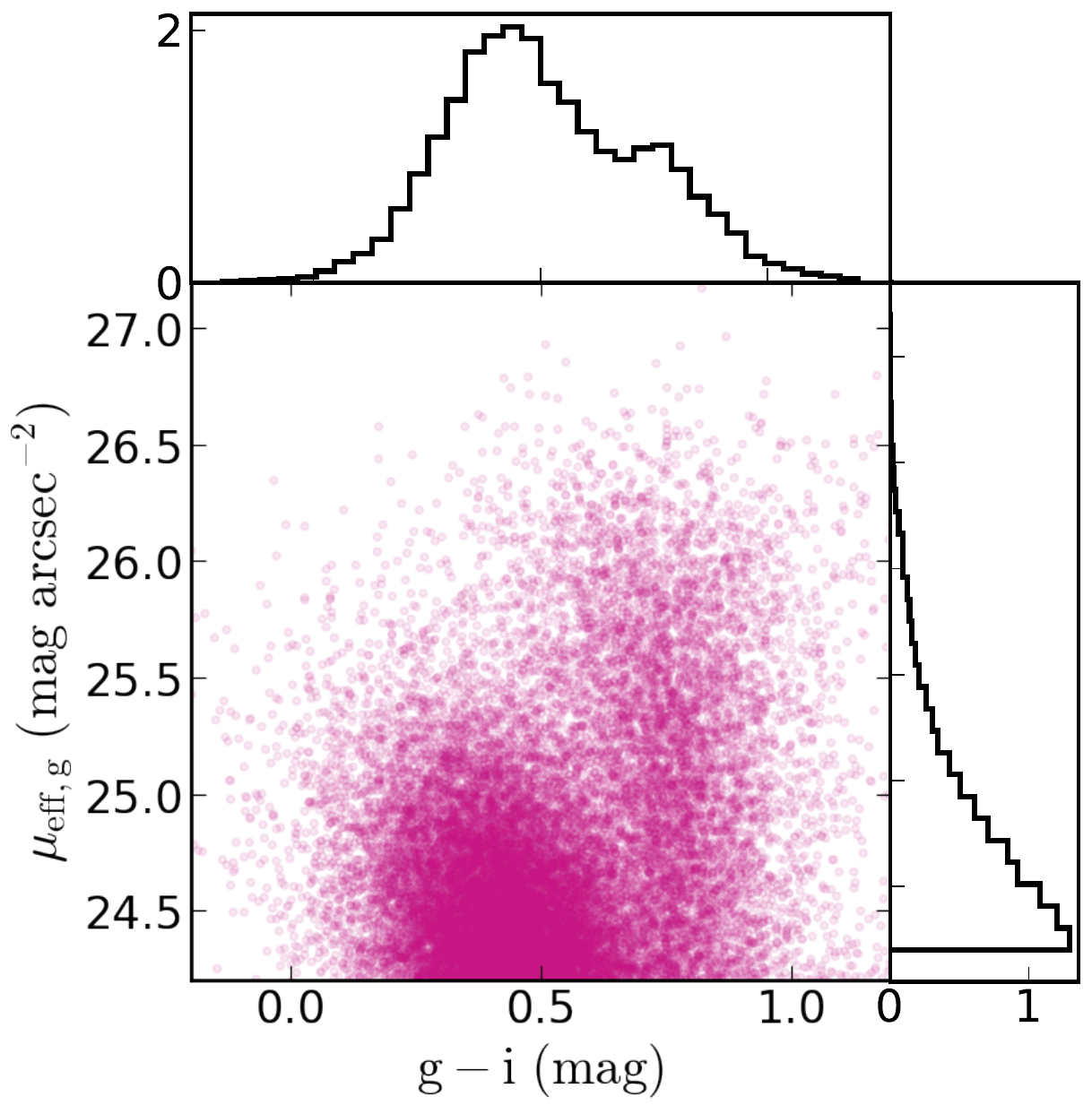}
\caption{ The complete sample of LSBGs from the Y3 DES catalog in color-surface-brightness space is shown along with the histograms along the two axes. Through visual inspection it is apparent that the red and blue LSBGs occupy different loci in this space.
\label{fig:col_sfb_space}}
\end{figure}

The color bimodality that is a feature of bright galaxies is also seen in the case of LSBGs, as indepedently demonstrated by \citet{2018ApJ...857..104G} and \citet{2021ApJS..252...18T}. They fit the $\gi$ color distributions for each of their samples with pairs of Gaussians and show that the ``red" and ``blue" sub-samples can be separated with appropriately chosen color thresholds. The values of the $\gi$ color thresholds are 0.64 and 0.60 respectively with the offset being the result of the DES sample being dominated by blue LSBGs. Furthermore, the blue LSBGs tend to be brighter in terms of their $\mug$ compared to the red LSBGs. We investigate this result further in the color-surface-brightness (CSB) distribution \citep{1994ApJS...93..397C,2022arXiv220502193Z} of our sample shown in Figure \ref{fig:col_sfb_space}. Here we search for patterns that can be associated with the stellar populations of the LSBGs. The one-dimensional distributions of each parameter are also shown in the top and right histogram. 

%Each point in this figure is scaled by the angular size of the LSBG in terms of $R_{\rm eff}$. 

The broadband $\gi$ color and $\mug$ represent two parameters that are explicitly distance independent at low redshifts, which motivates us to explore their joint parameter space. However we note that the $\gi$ color can be affected by extinction while completeness of the sample is limited towards large $\mug$. As seen in Figure 7a in \citet{2021ApJS..252...18T}, the angular size of the red and blue LSBGs appear to be distributed similarly. However we also note that the red LSBGs are more common in the extended lower surface brightness region. This appears as a digression from the absence of a correlation between color and surface brightness of LSBGs noted by \citet{1997PASP..109..745B}. According to \citet{2021ApJS..252...18T} this can be alievated if the size-luminosity relations for red and blue galaxies in SDSS \citep{2003MNRAS.343..978S} are extrapolated to low luminosities. This would mean that the red galaxies are larger, resulting in a lower surface brightness than their blue counterparts.

What we observe in Figure \ref{fig:col_sfb_space} is in agreement with this picture. There is a clear ``red sequence" made up of LSBGs with quiescent stellar populations that stretches down to the surface brightness limit of the survey. Alongside, there exists a corresponding star-forming concentration of brighter LSBGs making up the ``blue cloud". Unlike the blue cloud, the red sequence is spread out over a large range of surface brightness. 

\section{Host Galaxy Selection and their Properties} \label{sec:zm_samp}

The host galaxies by virtue of their mass exert a dominating influence on the dwarf galaxies around them. Therefore to understand how they quench the nearby LSBG population we begin by carefully selecting a sample based on their mass and environment and then define the boundaries of their halos. We associate LSBGs to them in projection, which makes it necessary to outline the mode of background subtraction that we apply. Here in this section we describe our methods in details.

\subsection{Sample}

We use the z = 0 Multiwavelength Galaxy Synthesis \citep[z0MGS][]{2019ApJS..244...24L} atlas of galaxies in the local universe (distances $D<100$ Mpc) to determine the host sample. This robustly determined catalog is based on ultraviolet, near-infrared, and mid-infrared images from NASA’s Wide-field Infrared Survey Explorer (WISE) and Galaxy Evolution Explorer (GALEX) missions. This catalog contains the stellar mass $\mathcal{M}_{\star}$, star formation rates (SFR) and distance estimates for the galaxies in this work. Their masses cover the LMC to MW mass range we are interested in, making this catalog befitting to use as the host sample in this work.

%The distances of the hosts are compiled from the existing Lyon Extragalactic Database (LEDA), Extragalactic Distance Database (EDD) and CosmicFlow databases that uses a variety of modeling, e.g., tip of the red giant method (TRGB), Hubble Flow, Tully-Fisher or group assignment. In light of the heteregenous nature of the distance estimates, different thresholds are put on the uncertainties while Hubble Flow based redshifts are used for $v > 3500$ km s$^{-1}$, such that the typical uncertainty is $\approx 0.1$ dex. The stellar masses are derived from the near-IR WISE1 intensity at 3.4$\mu$m using the mass to light ratio $\Upsilon_{*}^{3.4}$ that is calibrated using the GALEX-SDSS-WISE Legacy Catalog \citep{2016ApJS..227....2S}. 

\subsection{Halo Boundary} \label{sec:halo_bound}

We define the boundary of the host halo as the virial radius, particularly adopting the definition where this corresponds to the extent within which the density of the halo is 200 times the critical density, $\mathcal{R}_{200}$. For this purpose, we need to calculate the virial mass $M_{200}$ first, which we derive from the stellar masses $\mathcal{M}_{\ast}$ provided in the \zm{} catalog and a stellar-halo mass relation (SHMR) relation. For this purpose we adopt the robust SHMR of \citet{2010ApJ...717..379B} that is established through abundance matching. We use the same definition of the halo boundary for the hosts residing in cluster as those in isolated environments.

Having defined the halo boundary, we proceed to define the zones around the halo that will be relevant in this study. Galaxies located at 3D radial distances within $\mathcal{R}_{200}$ are typically considered to be satellite galaxies. The associated population of galaxies we are interested in are located beyond $\mathcal{R}_{200}$ . In this work we follow the prescriptions of \citet{2019MNRAS.483.1314B,2021ApJ...909..112D} and adopt $2.5\mathcal{R}_{200}$ as the outer boundary of the associated region. This limit is appropriate as the properties of the associated galaxies change significantly beyond this extent.

%While the outer boundary can be taken to be the splashback radius \citep{2015ApJ...810...36M} which for a MW sized halo would be $\approx 2R_{200}$,

\subsection{Host-LSBG association}

%While it is known that HSBGs tend to cluster close to massive hosts \citep{2014ApJ...788..188S,2020ApJ...902..124C,2021MNRAS.503.4976X}, a similar trend has been noticed for LSBGs as well \citep{2021ApJS..252...18T,2022arXiv220411883E}.

In order to connect the LSBGs as having interacted with the \zm{} hosts, we require the distance information for the former to match that of the latter. While the distances to the hosts are known, this does not follow for the LSBGs because the dataset we use is based on photometry alone. The classical method of spectroscopically measuring distances (e.g. \citet{2017ApJ...847....4G}) and ascribing satellites, backsplash or infalling galaxies around a host becomes prohibitively expensive in observing time when we focus on the LSBGs among them \citep[e.g.,][]{2021ApJ...923..257K,2023arXiv230300774G}.  The surface brightness fluctuation (SBF) method \citep{2019ApJ...879...13C} has been applied for the purpose of measuring distances to photometry derived catalogs of LSBGs \citep{2019ApJ...879...13C,2022arXiv221100629C}. However, this method can at best be used to probe distances up to a few Mpc with ground-based observations like the DES \citep{2021ApJ...908...24G}. Given the nature of LSBGs it is time intensive to observe a large sample as well. Therefore, these qualities of our sample render the application of the SBF method unsuitable for this work. 

Instead, we use the simple yet commonly used method of associating the LSBGs with the nearest host based on their projected angular separation \citep{2015ApJ...798L..45V,2011ApJ...731...44N,2014MNRAS.442.1363W,2016A&A...590A..20V,2022arXiv220502193Z,2022arXiv221014994L}. However, this method introduces contamination from background galaxies that needs to be statistically subtracted. We describe this process in the following section.

%\textbf{include \citep{2022ApJ...927..121W,2005ApJ...618..557N}?}

\subsection{Background subtraction}

When viewing the distribution of LSBGs associated with hosts in projection on the sky, contamination may arise from what are known as interlopers$- $ galaxies in the foreground or background of the host that appear proximate in projection \citep{1992ApJ...400...74Z}. When estimating the level of this contribution, while doing so ``globally" is simpler, this leads to biases arising from the way dark matter structure in the universe has been hierachically formed \citep{1978MNRAS.183..341W} and means that there will be larger numbers of interlopers in and around hosts in high density environments and vice versa. A way to circumvent this is to consider the contamination ``locally" \citep{2006ApJ...647...86C,2014MNRAS.442.1363W,2020MNRAS.496.5463A,2021MNRAS.505.5370T}, wherein the contamination signal is evaluated from annular regions around the hosts in consideration. 

In the subsequent analysis, we choose annular regions corresponding to separations of $2.5 <\theta/ \theta_{200} < 4.5$ around the hosts to estimate the projected background density of LSBGs. For a given host, this corresponds to a projected angular area $14\times$ and $2.24\times$ larger than the halo and associated regions. For a constant background density of LSBGs, this implies a $S/N$ of 3.46 and 1.73 respectively. A larger area implies that the background estimate is robust to Poisson fluctuations.

Furthermore, since the \zm{} hosts are selected on the basis of a cut on their projected virial radii $\theta_{200}$, this reduces the background contamination. This is because, given a background density of LSBGs, the number of contaminants would increase with a host subtending a large area on the sky.

\subsection{Environment} \label{sec:fof_algo}

\begin{figure*}
\centering
\includegraphics[scale=0.425,trim={0 0.3cm 0 0.2cm},clip]{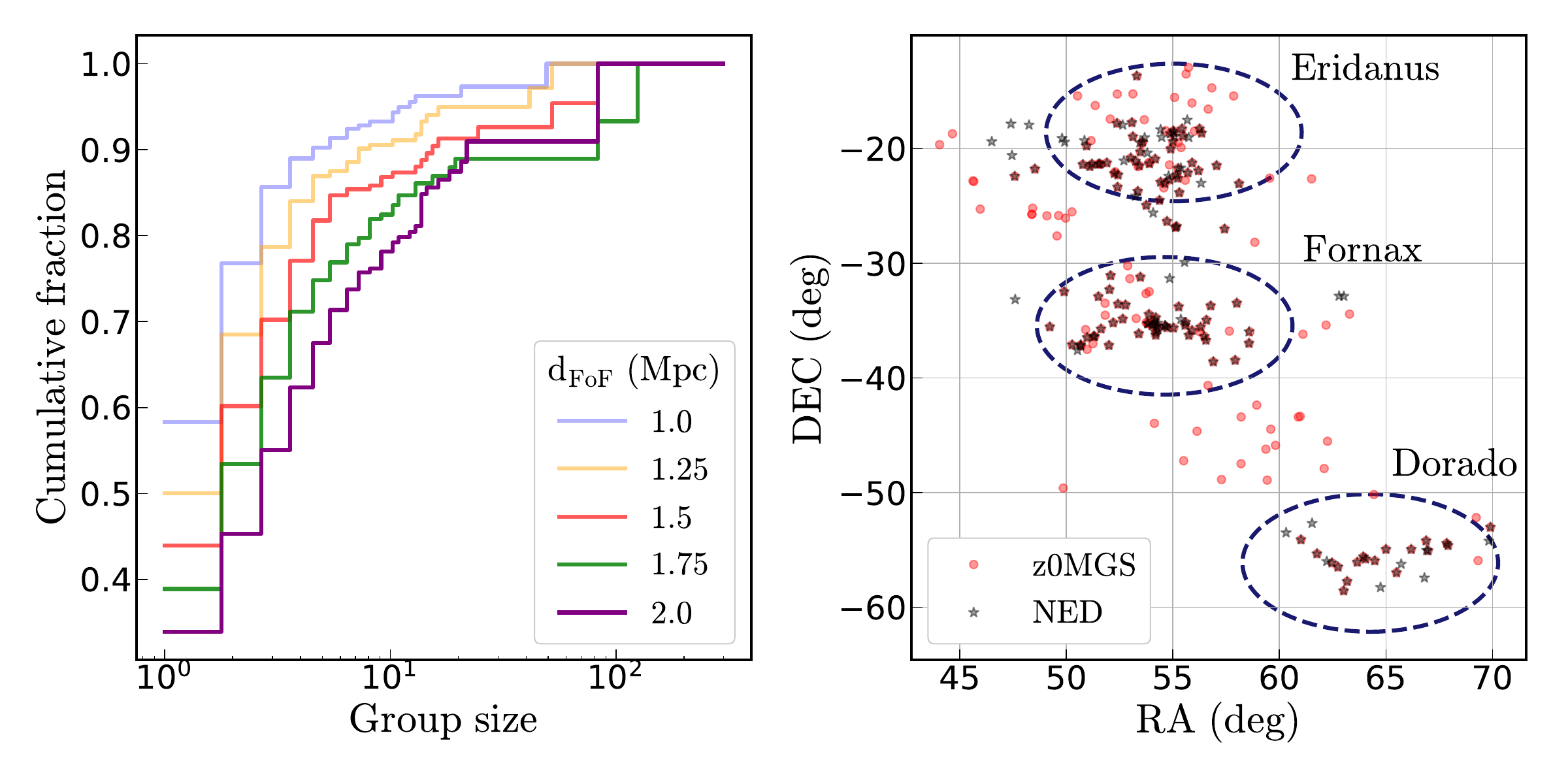}
\caption{\textit{Left:} The cumulative fraction of the \zm{} hosts in the DES footprint as a function of their association size for different choices of the linking length $1.0 <d_{\rm FoF} < 2.0$ Mpc. We choose $d_{\rm FoF}=1.5$ Mpc so that our isolated hosts are robust in terms of their isolation criteria and the clusters hosts are complete in the Fornax region as well. \textit{Right:} Map of the Fornax region showing the Fornax, Eridanus and Dorado overdensities of galaxies. The \textit{red points} represent galaxies from the \zm{} catalog making up the four largest FoF groups when using $d_{\rm FoF} = 1.5$ Mpc. We compared this with bright galaxies from NED in the same region that have been shown using the \textit{gray stars}. This shows that our sample of hosts in the cluster environment is complete. \label{fig:fof}}
\end{figure*}

The all-sky \zm{} catalog comprises 15748 members of which 2034 lie within the DES footprint. We employ a Friends Of-Friends (FoF) algorithm to identify regions in the DES footprint with an overdensity of \zm{} galaxies and thereby classify them and the associated LSBGs according to their local environment. FoF algorithms are frequently useful in identifying overdensities in the universe and halo finding in simulations \citep[e.g.][]{1985ApJ...292..371D,2013ApJ...762..109B}. In the context of our work, the FoF algorithm is appropriate in separating out the hosts in the high and low density environments. Compared to works like \citet{2007ApJ...671..153Y} which sought to create group catalogs, our application of the FoF algorithm is simpler.
For this purpose we utilise the RA, DEC and distance estimates from the host catalog to map out the 3D positions of the galaxies using the \texttt{astropy.coordinates} package. Following this with an appropriately chosen linking length $d_{\rm FoF}$, we identify groups of spatially connected hosts. 

Primarily, we want to obtain two samples out of this exercise$- $ a ``cluster" sample corresponding to the hosts in the Fornax region and another ``isolated" sample which are hosts dwelling in the low-density field. Using Figure \ref{fig:fof} we lay out the reasoning behind the value of $d_{\rm FoF}$ used in this work. In the left inset, we plot the cumulative fraction of the \zm{} hosts in the DES footprint as a function of size of the FoF group. The different colored lines show this fraction for various values of $d_{\rm FoF}$ chosen between 1-2 Mpc. This was done to determine the best choice of $d_{\rm FoF}$ with the constraints that this would ensure completeness of the host sample in the Fornax region while keeping the isolated host sample remote from higher density environment. Those FoF groups that were identified as containing single galaxies made up the isolated sample whereas the largest few groups of $\sim 100$ member galaxies were found to be located near the Fornax region. We notice that the primary effect of taking a smaller value of $d_{\rm FoF}$ is to increase the number of isolated hosts and decrease the number of the hosts in the cluster environment. However, we are unable to indefinitely increase the value of $d_{\rm FoF}$ because it tends to an incomplete identification of the cluster hosts.

We attain a balance by taking $d_{\rm FoF} = 1.5$ Mpc which is reasonable because it satifies the constraints for both cluster and isolated samples. Out of 2034 galaxies$- $ 195 are in the cluster environment corresponding to the four largest FoF groups, 894 are isolated, and 945 are associated with intermediate sized FoF groups. In the right panel of Figure \ref{fig:fof} we show the map of the cluster hosts that we identified using $d_{\rm FoF} = 1.5$ Mpc and compare them with member galaxies in the same region obtained from the NASA Extragalactic Database (NED). We find that our cluster sample is reasonably complete around the Fornax, Eridanus and Doradus overdensities and takes into account those galaxies located in the filaments between them. Our choice is consistent with the definition adopted in \citep{2012ApJ...757...85G,2021ApJ...915...53D} wherein galaxies at separations $> 1.5$ Mpc are considered isolated. Such a choice of $d_{\rm FoF} = 1.5$ Mpc is more conservative than other examples in the literature (e.g. \citet{2014MNRAS.442.1363W,2020ApJ...898L..15B}) but it nonetheless enables us to robustly select the sample of isolated hosts.

%This is highlighted through the right panel of Figure \ref{fig:fof} where we plot the ratio of the linking length to the virial radius $\mathcal{R}_{200}$ of the hosts $d_{\rm FoF}/ \mathcal{R}_{200}$ as function of the host stellar mass $\mathcal{M}_{\star}$. For this purpose, the SHMR used in \citet{2010ApJ...717..379B} is used. Among the curves for the various $d_{\rm FoF}$ used, we find that more massive galaxies tend to have virial spheres extending closer to each other. The dashed line represent $d_{\rm FoF} = 4.5 \mathcal{R}_{200}$. For the most massive hosts in our sample ${\rm log}(\mathcal{M}_{\star}/ M_{\odot}) > 10.5$, the ratio is above this value.

\begin{figure*}
\centering
\includegraphics[scale=0.35,clip]{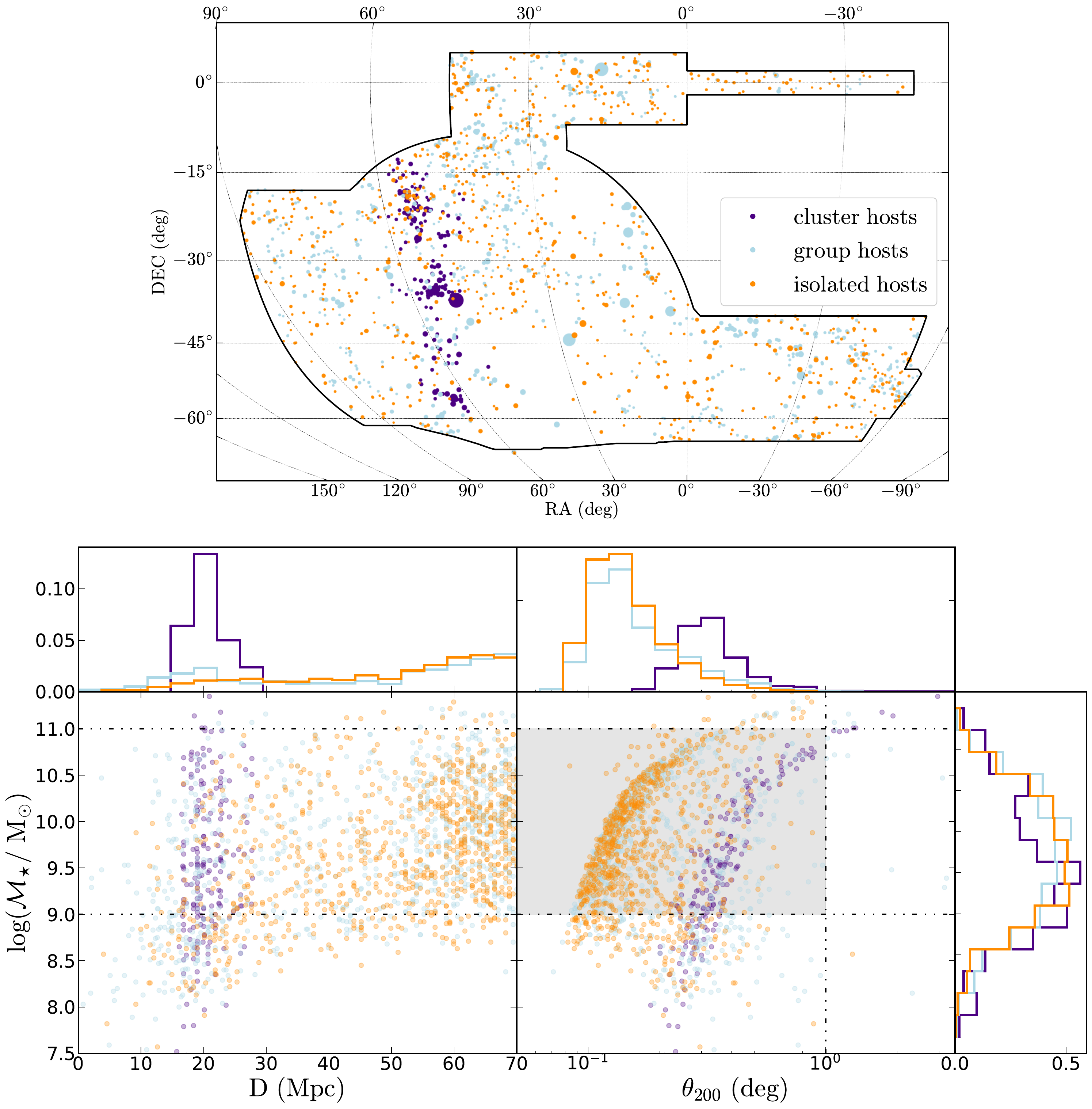}
\caption{ \textit{Upper panel:} The distribution of \zm{} galaxies in the DES footprint with the colors \textit{indigo, light blue} and \textit{orange} depicting those belonging to the cluster, groups and isolated subsamples by the FoF algorithm respectively. The size of each point is proportional to the projected virial radii $\theta_{200}$ of the respective galaxy. \textit{Lower left panel:} The distribution of these galaxies in $\mathcal{M}_{\star}- D$ stellar mass-distance space. The horizontal \textit{dash-dotted} lines show the mass cuts that have been adopted in this work $9.0<{\rm log}(\mathcal{M}_{\star}/M_{\odot})<11.0$ with the \textit{shaded gray} box representing the region of the space from which our sample of hosts are drawn. \textit{Lower right panel:} The distribution of these galaxies in $\mathcal{M}_{\star}- \theta_{200}$ stellar mass-projected virial radius space. The vertical \textit{dash-dotted} line is the cut on $\theta_{200}=1.0$ deg that we have adopted here. The histograms show the marginalized distributions for each of the axes. \label{fig:host_prop_space}}
\end{figure*}

The respective host samples have been mapped out over the full DES footprint in the upper panel of Fig. \ref{fig:host_prop_space} whereas the distributions of stellar mass $\mathcal{M}_{\star}$, distance $D$ and projected virial radius $\theta_{200}$ has been plotted in the lower panel of the same.
The cluster and the isolated hosts sample have been depicted using indigo and orange colors respectively. Also shown are hosts in intermediate size groups in light blue color although we do not explore their effect on LSBGs further in this work. The cluster hosts in and around Fornax are prominent on the map as well as in the distance histograms. Fornax is the second closest cluster and has been the subject of thorough searches for dwarf galaxies \citep{1987AJ.....94.1126C,1989AJ.....98..367F}, with a known presence of a large number of LSBGs \citep{1988MNRAS.232..239D,1987AJ.....94.1126C,2015ApJ...813L..15M,2017A&A...608A.142V}. Associated with it is the Eridanus cluster \citep{1993ApJ...403...37G,2006MNRAS.369.1351B}. This association together with Dorado group is part of the larger Southern Supercluster \citep{1953AJ.....58...30D,1989AJ.....98.1175M} and is identified by the FoF algorithm as a single region. This provides a standard of reference to compare with the LSBGs that dwell in the low density regions of the DES footprint. 

%Since we are interested in the properties of the LSBGs that reside in high density environments, there is a significant benefit of this structure falling within the DES footprint as the processing of galaxies residing in this region have been well studied.

These structures together contain the largest number of \zm{} hosts in the footprint and correspondingly an overdensity of LSBGs. In \citet{2021ApJS..252...18T}, while Fornax comes up as two peaks in overdensity associated with Abell S373 located at $18.97 \pm 1.33$ Mpc, the Eridanus group is separately designated as the source RXC J0340.1-1835 and is located at a distance of $23.41 \pm 1.64$ Mpc (NED).

The \zm{} galaxies that make up the isolated host sample are approximately uniformly distributed across the DES footprint except at the locations of a few voids. Given the criterion for the FoF search used to select these hosts, this means that the nearest massive galaxy is beyond 1.5 Mpc away. This implies that this selection of host galaxies is more isolated in comparision to the MW, M31, M81, IC 342, Maffei 1 and Sculptor in the Local Volume \citep{2005AJ....129..178K}.

%More so, this means that our isolation criteria is comparable to that deployed by \citet{2012ApJ...757...85G}.

%Although we are able to identify hosts in groups with a few to few tens of members using the same algorithm, we do not use them in this work. This is because we are mainly interested in comparing the effect of the environments on LSBGs

\subsection{Sample cuts}

In order to restrict our sample to host galaxies with masses similar to the LMC ($2.7 \times 10^9 M_{\odot}$; \citet{2002AJ....124.2639V}) and the Milky Way ($(5.43 \pm 0.57) \times 10^{10} M_{\odot}$; \citet{2017MNRAS.465...76M}, $(6.08 \pm 1.14) \times 10^{10} M_{\odot}$; \citet{2015ApJ...806...96L}) we place cuts on the stellar mass such that 10$^9 M_{\odot} < \mathcal{M}_{\star} <$ 10$^{11} M_{\odot}$. We considered using the distance $D$ of these galaxies to restrict the sample of hosts to those situated in the low redshift. The \zm{} sample is incomplete at distances beyond 70 Mpc. Closer by, e.g. within 5 Mpc, the LSBGs are found to be shredded by virtue of the detection pipeline \citep{2021ApJS..252...18T}.

We instead rely on a cut on the projected virial radius $\theta_{200}$ of the host since reducing the level of background contamination is our main concern. The $\theta_{200}$ is calculated from the host stellar mass using the stellar-halo mass relation (SHMR) in \citet{2010ApJ...717..379B} to obtain the physical virial radius projected at the distance $D$. Choosing hosts with a small $\theta_{200}$ lets us mitigate the contamination from foreground/background LSBGs associated with not only nearby hosts projecting a large virial radius but also the distant hosts.

We choose to use an upper limit on the projected virial radius of these hosts $\theta_{200}<1.0$ deg to limit contamination from the background. This limits our sample to 856 host galaxies$- $ 138 in clusters and 718 in isolation. In Fig. \ref{fig:host_prop_space} we map the host sample selected on the basis of their environment from the \zm{} catalog in the upper panel. We show the distributions in $\mathcal{M}_{\star}- D$ space as well as in $\mathcal{M}_{\star}- \theta_{200}$ space in the lower panels, respectively

Adopting this cut on $\theta_{200}$ can potentially lead to a selection bias against the more massive cluster hosts, as seen in the right panel of Fig. \ref{fig:host_prop_space}. We see that the mode of the $\theta_{200}$ distribution, the cluster hosts have, by virtue of their proximity to us, is larger than the isolated hosts. Therefore imposing the upper limit on $\theta_{200}$ has the result of filtering out a few of the cluster hosts that have $\mathcal{M}_{\star} > 10^{10.5} M_{\ast}$.

\section{Identification \& Characterization of Quenched LSBGs} \label{sec:analysis}

Having classified the \zm{} hosts according to their environment, we seek to identify and characterize LSBGs around them. We undertake a novel attempt to identify satellite and associated LSBGs around MW and LMC mass isolated hosts and ascertain if quenching has affected them using their color and surface brightness information. The larger goal here being to identify the environmental dependence of the quenching mechanisms$- $ how they might vary between the extremities of cluster and isolated environments. This is a novel investigation of how quenching operates in associated LSBGs. For this purpose we use two distinct yet somewhat parallel methods$- $ the projected surface density profiles around the hosts and the distributions in color surface-brightness (CSB) space.

\subsection{Statistical signal of Satellite \& Associated galaxies} \label{sec:density}

A way to directly study overdensities of galaxies around massive hosts is to look at their radial distribution through projected surface density profiles \citep[e.g.][]{2005ApJ...618..557N,2005MNRAS.356.1233V,2014MNRAS.442.1363W}. This motivated \citet{2021ApJS..252...18T} to look at the radial profiles of LSBGs around density peaks in the DES footprint and which they found are comparable with the radial profiles for bright galaxies in the 2MPZ survey. Using radial profiles, Roberts et al. (in prep.) found overdensities around local universe hosts with LMC and MW masses. We now carry forward this investigation by looking at the projected density profiles of the LSBGs around the more extensive \zm{} host sample.

In order to determine the raw projected density profile, we first find the raw number counts of LSBGs around each host in bins of the angular separation normalized with respect to the projected virial radius,  $\theta/ \theta_{200}$. The bins cover the range $(0.0,2.5)$. The counts are then normalized with respect to the dimensionless annular area corresponding to the $\theta/ \theta_{200}$ of each bin. The choice of using this is motivated by the fact that different hosts have different masses and are situated at different distances, therefore working in dimensionless units enables us to regularize these host-to-host variations. We determine the background contribution to the density profile for each host by selecting an annulus with $2.5 < \theta/ \theta_{200} < 4.5$ and finding the average density of LSBGs in it. We subtract the background density from the raw projected density profile to determine the projected density profile  $\Sigma(\theta/ \theta_{200})$. We then stack these profiles and calculate the mean and the standard deviation using the bootstrap method wherein we sample among the hosts in 5000 iterations. The standard error on the mean is derived by appropriately normalizing by the square-root of the total number of hosts. 

\begin{figure*}
\centering
\includegraphics[scale=0.425]{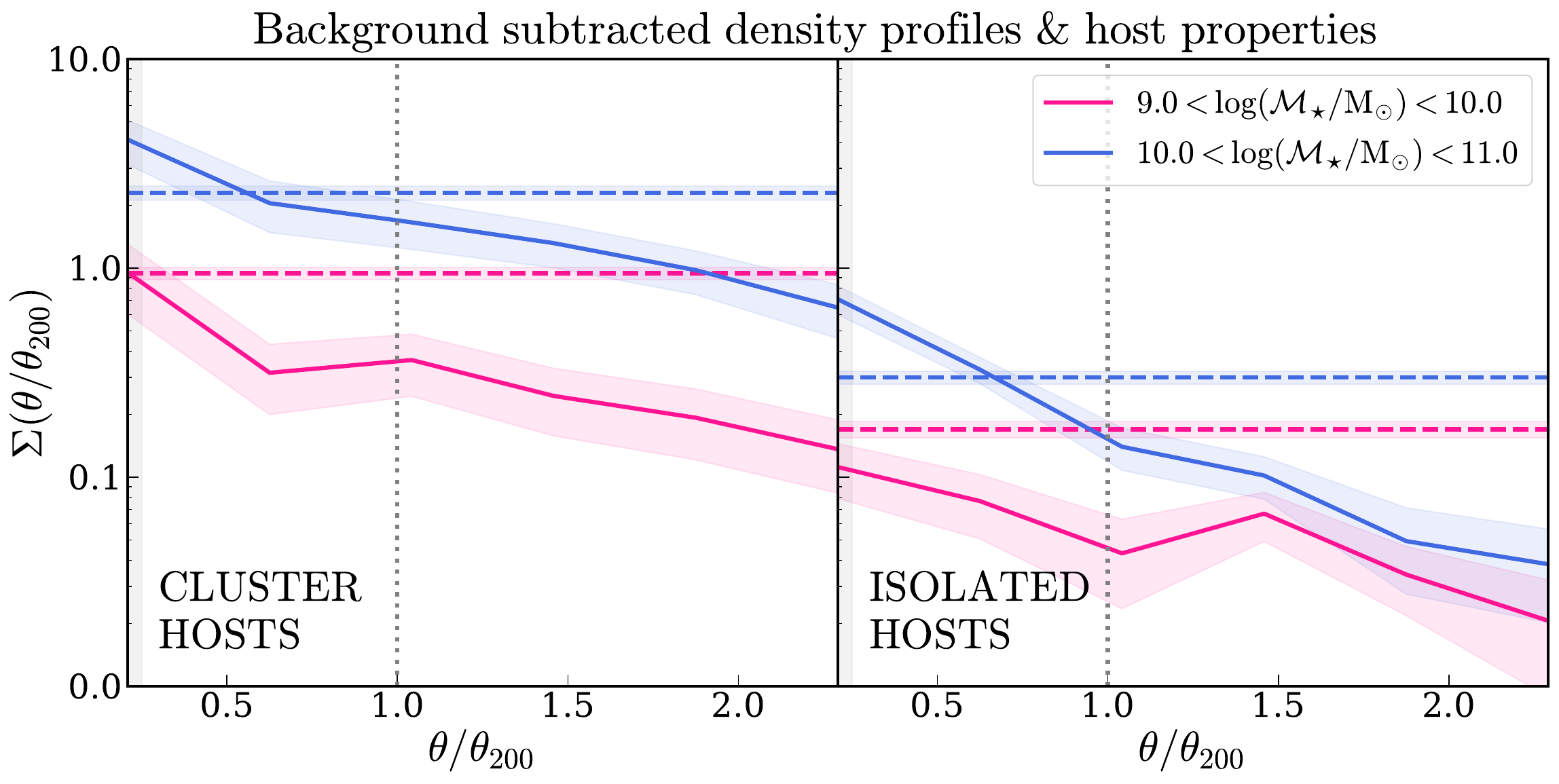}
\caption{The background subtracted projected surface density profile $\Sigma(\theta/ \theta_{200})$ of LSBGs around \zm{} hosts belonging to cluster (\textit{left panel}) and isolated environments (\textit{right panel}) with LMC like stellar masses $9.0<{\rm log}(\mathcal{M}_{\star}/M_{\odot})<10.0$ (\textit{pink line}) and MW-like masses ($10.0<{\rm log}(\mathcal{M}_{\star}/M_{\odot})<11.0$) (\textit{blue line}). The dashed horizontal lines show the average density of LSBGs at separations of $2.5 < \theta/ \theta_{200} < 4.5$ from the respective sample of hosts, which constitute a measurement of the background level. The colored shaded regions show the uncertainties on the $\Sigma(\theta/ \theta_{200})$ estimated using the bootstrap method. The dotted black line is $\theta/ \theta_{200}=1$ represents the extent of the typical virial radii of the hosts.  The \textit{gray} shaded region at $\theta/ \theta_{200} <0.25$ shows the region where incompleteness of the LSBG sample might arise due to blending with the host galaxies. \label{fig:radial_dist}}
%\caption{\textit{Top:}  The dashed profile represents the background density of LSBGs. The corresponding shaded regions are the bootstrap uncertainties on the profile.  \textit{Bottom:} The background subtracted projected surface density profile of LSBGs.  \label{fig:radial_dist}}
\end{figure*}

In Figure \ref{fig:radial_dist} we plot the density profiles $\Sigma(\theta/ \theta_{200})$ according the environment of the \zm{} host. The dashed lines show the level of background density that was subtracted to obtain the profiles. All the shaded regions signify the standard error from the bootstrap method. The extent of the profile for $\theta/ \theta_{200} <1$ and $1<\theta/ \theta_{200} <2.5$ correspond to the satellite and associated populations, respectively. Since there is possible incompleteness at small radial distances due to blending with the host galaxy \citep{2016A&A...590A..20V,2023arXiv230214108L} we highlight a conservative approximation of this domain at $\theta/ \theta_{200} < 0.25$ to show where this effect might be prominent. We performed a visual inspection and found blending effects were limited within this region.

%Since we are interested in the clustering of LSBGs around \zm{} hosts with stellar masses in the ranges of the LMC and MW, we separate out the population of all hosts across all environments into two bins of $9.0<{\rm log}(M_{\star}/M_{\odot})<10.0$ and $10.0<{\rm log}(\mathcal{M}_{\star}/M_{\odot})<11.0$ roughly corresponding to the measured stellar mass of each of the objects \citep{2002AJ....124.2639V,2017MNRAS.465...76M}. The solid lines in the left panel of Fig. \ref{fig:radial_dist}, show a concentration of LSBGs inside the virial radius $x <1$ which is more pronounced for the more massive hosts. On the right panel of Fig. \ref{fig:radial_dist}, the density profiles are shown with the hosts across all stellar masses divided according to their environment$- $ cluster or isolated, the details of which was provided in the previous section. We validate that the cluster environment is richer in LSBGs compared to the low density environment. However the gradients of these profile for $x>1$ are similar to each other.

Since we are interested in the clustering of LSBGs around \zm{} hosts with stellar masses in the range of the LMC and MW, we separate the host sample into two bins of $9.0<{\rm log}(\mathcal{M}_{\star}/M_{\odot})<10.0$ and $10.0<{\rm log}(\mathcal{M}_{\star}/M_{\odot})<11.0$ roughly corresponding to the measured stellar mass of each of the objects \citep{2002AJ....124.2639V,2017MNRAS.465...76M}. The corresponding profiles are depicted using the pink and blue colored lines in Figure \ref{fig:radial_dist} with the left and right panels depicting the profiles around hosts in the cluster and isolated environments respectively. We find that the radial profiles in the two mass bins closely resemble each other in both environments with a notable concentration of LSBGs inside the virial radius ($\theta/ \theta_{200}< 1$). However there is an enhancement in normalization of the profiles as well as the background level for the cluster hosts compared to the isolated hosts which is expected given the high density environment of the former sample. 

We tested the statistical significance of our result by comparing the raw number of satellite and associated LSBGs $N_{\rm LSBG}(\theta/ \theta_{200})$ with that of mock background LSBGs in the different bins of $\theta/ \theta_{200}$. For this we took the projected background density from $2.5 < \theta/ \theta_{200} < 4.5$ and obtained the expected number of mock LSBGs for each annuli of $\theta/ \theta_{200}$. This is used as a parameter for a Poisson distribution from which we sample the mock number of LSBGs $N_{\rm mock}(\theta/ \theta_{200})$. We calculated 10000 realizations and none produced more counts of $N_{\rm mock}(\theta/ \theta_{200})$ than the measured $N_{\rm LSBG}(\theta/ \theta_{200})$ except the largest radial bin for LMC analogous hosts in the isolated environment. In those cases the fraction of realizations where the mocks exceeded the measured count was $\sim 0.0012$. This shows that the detection of the overdensities of satellite and associated LSBGs near the \zm{} hosts is robust against contamination from the background.

The phenomena we notice here, when the profiles are differentiated by the host properties, is essentially representative of the same result of hierarchical structure formation \citep{1991ApJ...379...52W}. This is namely that massive halos and halos in denser regions collapse earlier, leading to richer substructure around the central galaxies harbored in them. Furthermore, the profiles smoothly decay beyond the virial radii of the hosts where the associated LSBGs are located. This shows that the quenching does not simply end at the virial radius and may continue beyond it, thus highlighting the presence of backsplash and infalling LSBGs around these hosts \citep{2009ApJ...692..931L}.

%The fact that we find the density is lower for LMC like hosts, has been shown in \citet{2022MNRAS.513.2673J} which studies the satellite population of similar hosts. More so, the presence of backsplash galaxies was found to be rarer around these less massive hosts.

\begin{figure*}
\includegraphics[scale=0.425]{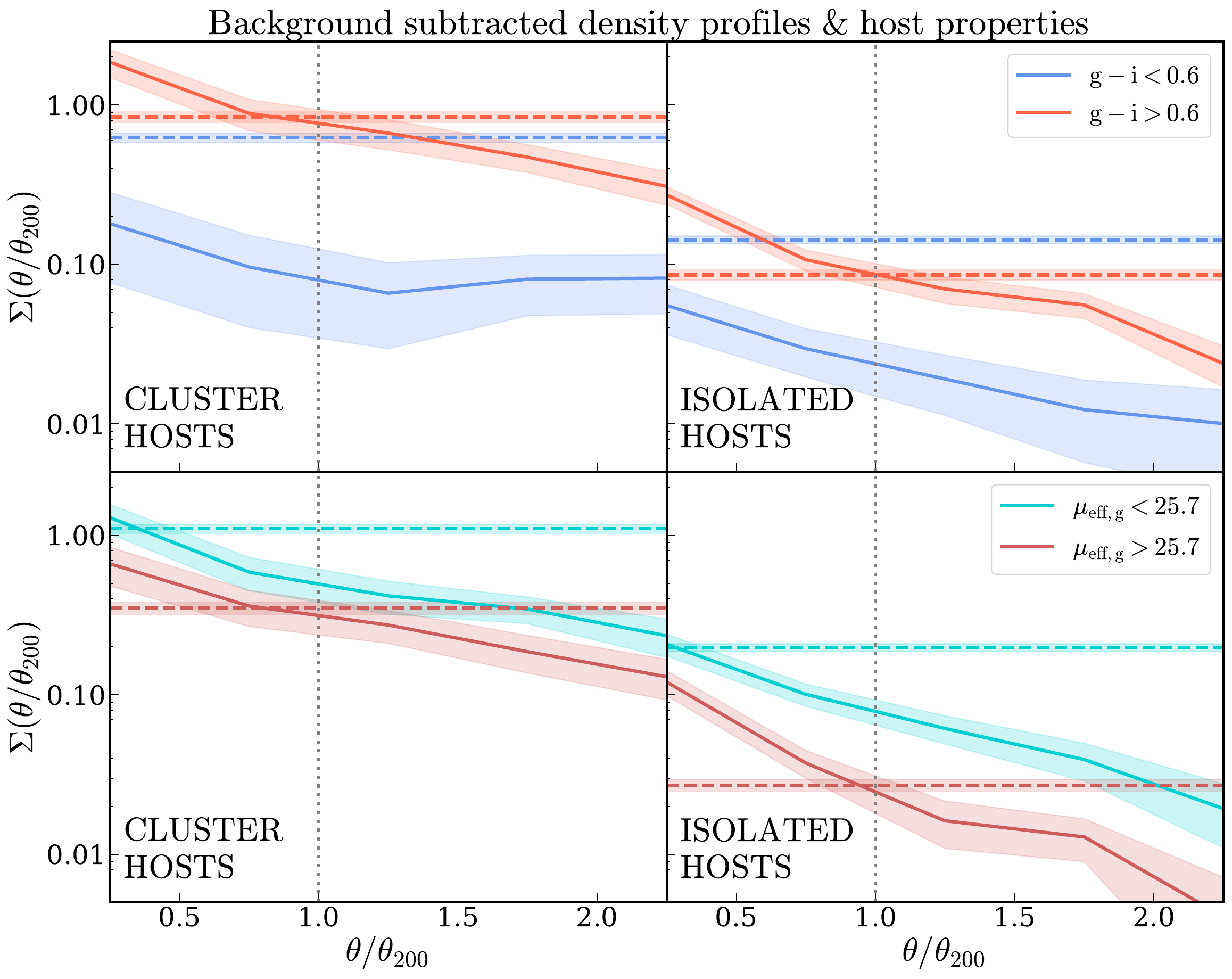}
\caption{The background subtracted projected surface density profile $\Sigma(\theta/ \theta_{200})$ of LSBGs around the \zm{} hosts belonging to cluster (\textit{left column}) and isolated environments (\textit{right column}) separated by the $\gi$ color of the LSBGs (\textit{upper row}) and by their surface brightness $\mug$ (\textit{lower row}). The dashed horizontal lines show the average density of LSBGs at separations of $2.5 < \theta/ \theta_{200} < 4.5$, which constitute a measurement of the background level. The colored, shaded regions show the uncertainties on $\Sigma(\theta/ \theta_{200})$ estimated using the bootstrap method.
The dotted black line is $\theta/ \theta_{200} = 1$ represents the extent of the typical virial radii of the hosts. The \textit{gray} shaded region at $\theta/ \theta_{200} <0.25$ shows the region where incompleteness of the LSBG sample might arise due to blending with the host galaxies. \label{fig:radial_color_sfb}}
\end{figure*}

We similarly obtain the background subtracted density profiles around hosts analogous to the MW in the mass bin $10.0<{\rm log}(\mathcal{M}_{\star}/M_{\odot})<11.0$, but this time we separate out the LSBGs by their $\gi$ and $\mug$ values. These profiles are shown in the upper and lower rows of panels in Fig. \ref{fig:radial_color_sfb} with the left and right columns showing the profiles around the cluster and isolated hosts, respectively. The LSBGs are separated into a "red" and "blue" populations by their $\gi$ color with a fiducial threshold value of 0.6 mag \citep{2021ApJS..252...18T}. When separating by their $\mug$ values into bright and faint populations, we use a threshold of $25.7 \,\sbfu$, which is the mean value of the surface brightness cuts that we employed (see Sec \ref{sec:lsbg_sample}).
%For comparision, the dashed horizontal lines show the level of background noise estimated from the average density in the region  $2.5 < \theta/ \theta_{200} < 4.5$.

We find that across both environments the profiles of red LSBGs ($g-i > 0.6$) have a higher concentration within the virial radius ($\theta/ \theta_{200}< 1$) relative to the background level. This occurs with an overall difference in normalization even beyond $\theta/ \theta_{200}< 1$ with respect to the profiles of its blue ($g-i < 0.6$) counterparts. The profiles for the cluster hosts are flatter and the background levels are also larger, especially for the red LSBGs. The profiles for the bright LSBGs ($\mu_{\rm eff,g} > 25.7\,\sbfu$) show a significant overdensity for $\theta/ \theta_{200}<1$, although they are lower in normalization compared to the fainter counterparts ($\mu_{\rm eff,g} < 25.7\,\sbfu$) in both types of host environments.

The presence of an overdensity of red LSBGs around these hosts is a signature of quenching of LSBGs taking place in the respective environments \citep{2022arXiv221003748K}. Noticeably, this is present in hosts in the clusters who themselves make up the substructure of the more massive cluster host as well as the isolated hosts in the low density environment. This shows the ubiquity of MW-like hosts ($10.0<{\rm log}(\mathcal{M}_{\star}/M_{\odot})<11.0$) in shaping LSBG evolution across diverse environments. In order to probe this further we consider the color-surface brightness distributions of satellites and associates separately in the next section.

%Since we are looking at the projected population of LSBGs around the hosts, it is expedient to compare our signal with the background density of LSBGs. To find this, for each host we select a random point in the DES footprint and repeat the analysis to derive the background projected density of LSBGs $\Sigma(x)_{back}$. This is depicted as the dashed lines in the upper panel of Fig. \ref{fig:radial_dist}, each of which shows a flat profile. In the lower panel, we subtract the background from the signal to give ``true'' projected density of LSBGs around the isolated hosts.

\subsection{Hess diagrams in Color$- $ Surface Brightness space} \label{sec:hess}

\begin{figure*}
\centering
\includegraphics[scale=0.325,clip]{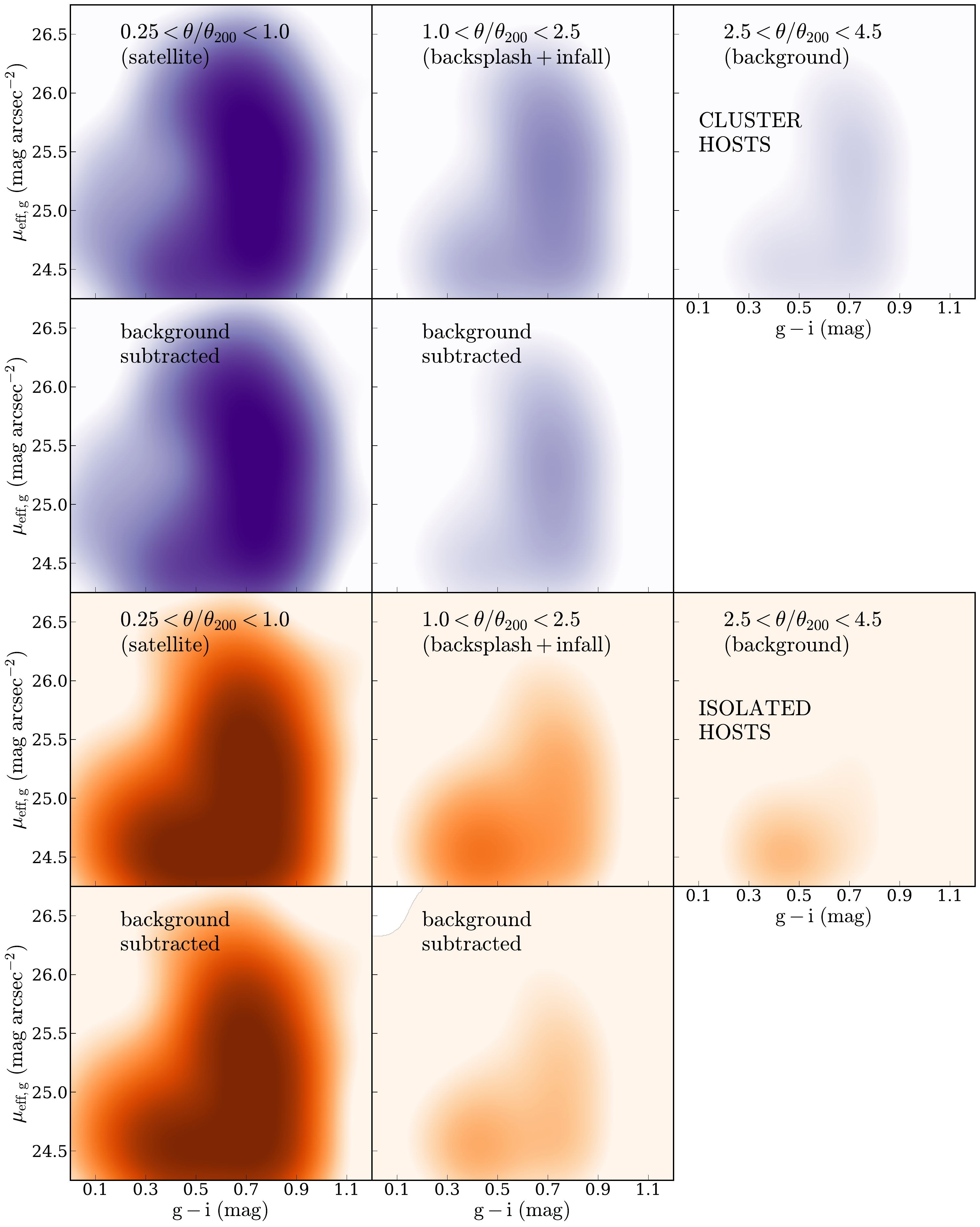}
\caption{The maps of Kernel Density Estimates (KDE) in the color-surface brightness (CSB) space of LSBGs around hosts with $9.0<{\rm log}(\mathcal{M}_{\star}/M_{\odot})<11.0$ belonging to the environments of clusters (\textit{top two rows, blue}) and in isolation (\textit{bottom two rows, orange}) and at different bins of angular separations $\theta$. The bins with $\theta/ \theta_{200} < 1$ (\textit{left}) and $1 < \theta/ \theta_{200} < 2.5$ (\textit{center}) corresponds to populations of satellites and associated galaxies (backsplash+infalling), respectively,  whereas the outer bin KDE with $2.5<\theta/ \theta_{200} < 4.5$ (\textit{right, first and third rows}) constitutes the basis for measuring the background contamination. The background subtracted KDE (\textit{second and fourth rows}) reveals a distinct red sequence visible for the satellites LSBGs in both cluster and isolated environments while the associated LSBGs in the latter show a less prominent red sequence and an enhanced blue cloud. These are the patterns of quenching taking place in the respective environments, while the associated LSBGs in the low density environment,  are seemingly still undergoing the process of quenching. \label{fig:quench_hess}}
\end{figure*}

%After we identify the color and surface-brightness of the LSBGs influencing their radial density profiles around the \zm{} hosts, w

We seek to explore the relationship between the photometric and spatial clustering properties further. Here we revisit the color-surface brightness (CSB) distribution of the LSBGs shown in Figure \ref{fig:col_sfb_space}. However with the insight gained from Section \ref{sec:density}, we look to incorporate the spatial density of these objects in their CSB distributions. Working with spatial densities as opposed to using raw counts, enables us to subtract the background contribution from the signal of interest in the CSB space.

For this purpose, the we employ the concept of a Hess diagram. Historically, they have been used to depict the spatial density of stars across regions of color-magnitude space to study globular clusters and dwarf galaxies \citep{Hess1924,1948AJ.....53..193G}. Although those studies focused on resolved stellar populations and especially the identification of the main sequence, it is essentially parallel to our investigation of the red sequence of quenched LSBGs in the CSB space. We start by determining the probability density of LSBGs in the CSB space. Instead of working with densities computed on discrete 2D bins, we choose the smoothness and continuity of distributions that is guaranteed by using a Kernel Density Estimator (KDE).  A Gaussian kernel with bandwidth of 0.1 is used for the KDE, after having renormalized the parameters to the interval $(0,1)$. We ensure that the bandwidth is larger than the typical uncertainty of a point in the CSB space. On the other hand, making the bandwidth arbitrarily large can dissolve features in the CSB distribution.

We show the KDE for the LSBGs in different environments using the Hess diagrams in Fig. \ref{fig:quench_hess}. The upper half correspond to the cluster and the lower half the isolated environments. For each of these hosts we look at nearby LSBGs and then split their population according to their projected separation from the host $\theta$ normalized by the projected virial radius of the host $\theta_{200}$. Three regions are $\theta/ \theta_{200} < 1$ for the satellites, $1 < \theta/ \theta_{200} < 2.5$ for the associated galaxies and we use $2.5 <\theta/ \theta_{200} < 4.5$ to estimate the contamination from background objects. The KDEs corresponding to these three regions are displayed in the three columns from left to right. We normalize the KDE with the total angular region of the footprint as considered in each bin. We then subtract the background KDE from the target KDE to give us the KDE shown in the right column.

Fig. \ref{fig:quench_hess} shows the background subtracted KDEs for the satellite and associated LSBGs across the two environments, with the darker regions representing a higher density of LSBGs at a given $\gi$ and $\mug$. The distributions belonging to the satellite and associated regions in the cluster environment show a distinct red sequence at $g-i > 0.6$ stretching across $24.5 < \mug <26.5\,\sbfu$ in both the satellite and associated regions. The distributions near the isolated hosts also show prominent red sequences for both regions. However it is at $\mug <25.0\,\sbfu$ where the distributions show an interesting deviation between the environments. We notice an enhanced blue cloud around $0.1 <g-i< 0.4$ in contrast to the cluster distribution. There is also an enhancement across $0.4 <g-i< 0.6$ between the two features that is the green valley \citep{2003MNRAS.341...54K}.

%The fact that we observe a red sequence in the distribution for the associated galaxies shows that, the probability of finding a quenched LSBG with $g-i> 0.6$ beyond the virial radius is similar among hosts residing in diverse environments.

\begin{figure*}
\centering
\includegraphics[scale=0.35,clip]{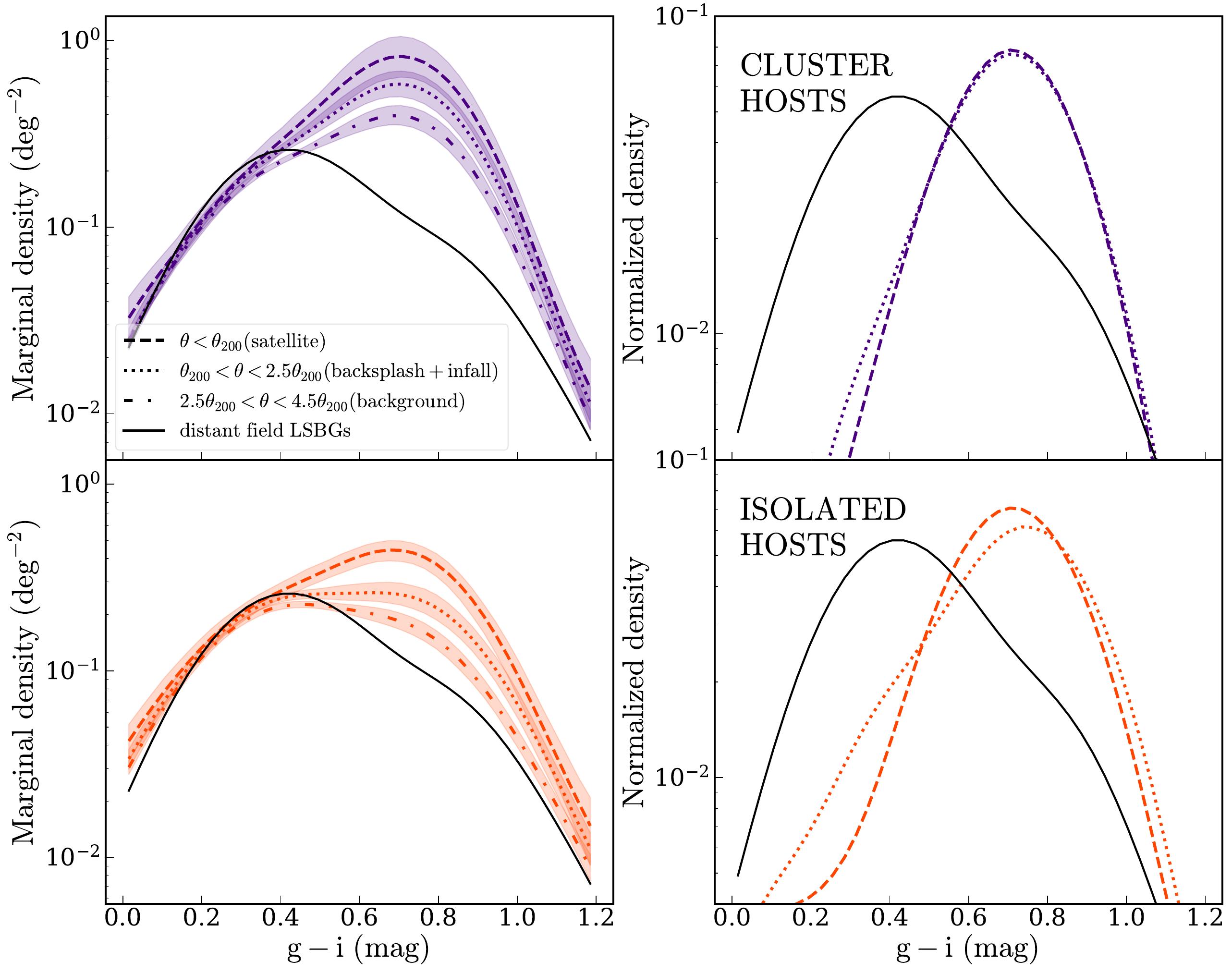}
\caption{\textit{Left:} The marginalized KDEs as functions of $\gi$ are shown for the satellite ($0.25 < \theta/ \theta_{200} < 1$), the associated ($1 < \theta/ \theta_{200} < 2.5$) as well as the background ($2.5 < \theta/ \theta_{200} < 4.5$) LSBGs as the \textit{dashed}, \textit{dotted}, and \textit{dash-dotted} lines respectively. The cluster and isolated environments of the LSBGs are coded using \textit{blue} and \textit{orange} colors respectively. The \textit{black} histogram is that of the distant field sample not within $2.5 \theta/ \theta_{200}$ of a \zm{} host in the footprint. \textit{Right:} The marginalized KDEs for the satellite and associated from which the background has been subtracted and re-normalized so that the areas under the curves are equal to one. All the color distributions of the satellite and associated galaxies deviate significantly from that of the field sample. They peak at $g-i = 0.72$ where there is a notable \textit{red excess} with respect to the distant field value. This shows that the LSBGs are being quenched by virtue of their environment, near the \zm{} hosts, rather than intrinsic reasons. \label{fig:quench_col_hist}}
\end{figure*}

\subsection{Marginalized Color distributions} \label{sec:gi_distrib}

To shed more light on the detection of the red sequence, we marginalize these distributions over the surface brightness axis to yield the $\gi$ color distributions for the LSBGs in the various regions and compare with the color distribution for a field sample of LSBGs. This lets us explore the color bimodality aspect of the LSBG distributions as seen in Fig. \ref{fig:col_sfb_space}. We plot these distributions in the left panel of Fig. \ref{fig:quench_col_hist} with the right panel showing the same distributions with background subtraction and normalization such that the area under the curves equals to the average number density of the respective sample. We depict both the satellite ($\theta/ \theta_{200} < 1$) as well as the associated ($1 < \theta/ \theta_{200} < 2.5$) populations as the dashed and dotted lines respectively. The shaded regions represents the uncertainties that have been determined through bootstrap resampling. The dash-dotted line in the left panel show the background ($2.5 < \theta/ \theta_{200} < 4.5$) distributions which is subtracted from the satellite and associated distributions. While the simple marginalized distributions on the left show the number densities of each LSBG population, the normalized distributions on the right enables us to compare their shapes.

The KDE for the $\gi$ distribution of the ``distant field" LSBG sample is shown as the black solid line. This sample is conservatively selected by ensuring that these LSBGs are not within 2.5$\theta_{200}$ of any \zm{} galaxy. To mitigate boundary effects we take into account the \zm{} members beyond the footprint as well during the process of selection. The equivalent area in the footprint that is not within 2.5$\theta_{200}$ of any \zm{} galaxy is 1565 deg$^2$ which we determined using a Monte Carlo method. The way the distant field sample is selected ensures that these LSBGs are situated in voids at low redshifts. In the latter case, we expect massive, luminous LSBGs to dominate unlike the faint, small LSBGs found near us \citep{2021ApJS..252...18T}.

We find that all of the LSBGs distributions in the cluster environment show broad peaks with a heavy tail towards the blue side. The distributions peak at $\gi = 0.72$. Concerning the LSBGs around the isolated hosts, we find that the peaks of the distributions are significantly more broad. While the satellites have a peak in the distribution close to where the cluster LSBGs peak, the associated LSBGs are more skewed. We find a difference between the satellites and associated LSBGs in both environments from the very blue tail of the distribution $g-i \gtrsim 0.0$. In contrast, the red tails at $\gi \gtrsim 1.0$ show similarity for the two populations in both cases. The distribution of these field LSBGs by inspection is bluer than that for the LSBGs near the \zm{} hosts. All the LSBG distributions show a significant excess near their peaks at $\gi = 0.72$ with respect to the field distribution. There is also a notable difference at the red and the blue ends of the distribution except for the case of associated LSBGs of isolated hosts showing a convergence at $\gi \gtrsim 0.4$.

We apply the two sample 1-D Kolmogorov–Smirnov test (K-S test) to test the dissimilarity of the different distributions. Comparing the distributions for the satellite and associated populations, we find a p-value that is $<0.01\%$ in both the cluster and isolated environments. We can therefore reject the null hypothesis that the satellite and associated populations are drawn from the same distribution. It is also interesting to compare these distributions with the field population of LSBGs. We find from the p-value which is $<0.01\%$ that the LSBG populations of satellites and associates near the \zm{} hosts show $\gi$ distributions distinct from each other as well as those selected from out in the field.

%The detection of a red sequence of galaxies in the Hess diagram followed by red excess in the color distribution, for all of the LSBGs located near the \zm{} is points towards quenching by the latter. Contrasting with the colors of a field sample shows that this form of quenching is unlike to be of intrinsic origin, e.g., reionization and supernova feedback. It is rather of environmental origin, where it is the massive galaxy influencing the gas and thereby the stellar content of these LSBGs. 

\section{Discussion} \label{sec:discuss}

In this work, we measure and characterize LSBG populations in the cluster and isolated environments they are embedded in. We place our key findings from Sec. \ref{sec:analysis} in context of the contemporary understanding of the field. Our findings can be summarized as follows:  1. We detect an overdensity of red and bright LSBGs in the radial density profiles near \zm{} hosts with a wide range of masses that are situated in both dense and sparse environments
 in Sec \ref{sec:density}. 2. We identify a red sequence of LSBGs in their color-surface brightness (CSB) space in Sec \ref{sec:hess}. 3. We find a red-excess with respect to the distant field population in the $\gi$ color distribution of \ref{sec:gi_distrib}. These are statistical detections made after correcting for background subtraction and point towards the host galaxies influencing the properties of the LSBGs beyond their virial radius. These results direct us to explore the importance of the connection between hosts and both backsplash and infalling galaxies in Sec. \ref{sec:backspl_disc}. Then in Sec. \ref{sec:quench_disc} we touch upon the theme of environmental quenching like pre-processing in low density environments that is an effect of this connection.
 
\subsection{Backsplash \& Infalling Galaxies} \label{sec:backspl_disc}

Outside the virial radius of a central galaxy, there exists a heterogeneous population of lower mass galaxies. This includes backsplash galaxies which have already completed pericentric passages as well as those on their first infall into the central. The former are expected to be quenched as result of the pericentric passage and to appear as older, redder stellar populations relative to the infalling galaxies and those in the field \citep{2014MNRAS.439.2687W,2018MNRAS.478..548S,2023MNRAS.tmp...27F}. Echoing \citet{2009ApJ...692..931L}, we collectively call these galaxies which are located at projected separations of $\theta_{200}< \theta <2.5\theta_{200}$, associated galaxies. The existence of such galaxies show that the spheres of influence of massive centrals extend far beyond their virial radii \citep[e.g.][]{2013MNRAS.430.3017B,2014ApJ...787..156B}. Using simulations \citet{2021MNRAS.501.5948B} and \citet{2022arXiv220510376B} show that out of all the associated galaxies beyond the virial radii of MW-like hosts, the fraction of backsplash galaxies is $\approx 50\%$ between $1-1.2 \mathcal{R}_{200}$. The ubiquitousness of these galaxies is pointed out by their presence in a diverse range of environment \citep{2000ApJ...540..113B,2009ApJ...697..247W,2018MNRAS.478..548S}. For example, the Tucana and Cetus galaxies \citep{2007MNRAS.379.1475S,2012MNRAS.426.1808T,2022arXiv220702229S} are dwarf backsplash candidates around the Local Group. 

LSBGs are known to constitute a sizeable subset of the low mass galaxies in the vicinity of MW and LMC mass hosts in the Local Volume \citep{2022arXiv221011070K}, including those hosts in isolated environments. LSBGs which are on backsplash and infalling orbits around massive central have been observed in simulations \citep[e.g.][]{2021ApJ...906...96A}, with backsplash galaxies occasionally found as far as $\approx 5\mathcal{R}_{200}$ from the more massive central
\citep{2012MNRAS.426.1808T,2021NatAs.tmp..160B}. In this work we not only detect LSBG populations around \zm{} hosts across different environments, including those with LMC like masses, we demonstrate that the population of red LSBGs extends to at least $2.5\times$ the projected virial radius. These populations are characterized by their relative redness and low surface brightness with respect to LSBGs in the field and indicate they have interacted with the hosts during the course of their backsplash or infall orbits. Our detection of red LSBGs around MW and LMC analogous hosts therefore establishes that LSBGs which also happen to be associated galaxies constitute an important component of the substructure of these hosts.  

Our findings highlight the importance of accounting for galaxies at the intersection of LSBGs and associated galaxies in order to gain a complete understanding of the outer extents of the halos of MW and LMC mass galaxies in isolated, group and cluster environments. There have already been challenges to the classical ``spherical overdensity'' based classical definition of the virial radius with the ``splashback'' definition \citep{2014ApJ...789....1D,2014JCAP...11..019A,2015ApJ...810...36M}. The latter represents the extent of dark matter material at their first apocenter around the host and also changes how we look at subhalos around the hosts \citep{2021ApJ...909..112D}. For example, if we defined the halo boundary by the splashback radius rather than the viral radius the associated galaxies considered in this work would now be classified as satellites instead. When galaxies around cluster mass hosts are selected according to their colors and their splashback radius measured from their density profiles \citep{2017ApJ...841...18B,2021ApJ...923...37A}, it is seen that red galaxies have been in the cluster longer compared to the blue galaxies. While the splashback radii of cluster scale systems have been studied extensively \citep{2016ApJ...825...39M}, the presence of associated galaxies enable a novel method to probe not only the splashback radii but the structure and composition of the outer halo beyond the virial radius of group scale or isolated systems.

In such associated galaxies, we get a better understanding if their HI gas distribution that is susceptible to removal through ram-pressure stripping \citep{2018MNRAS.478..548S} can be traced. Dwarf galaxies within the Local Group are known to be deficient in their HI content \citep{2012MNRAS.426.1808T,2021ApJ...913...53P}. Ever though it is difficult to constrain the HI content of LSBGs given current instrumental sensitivities \citep{2022MNRAS.516.1781Z}, there is promise for the future with the upcoming WALLABY \citep{2020Ap&SS.365..118K} and SKA \citep{2009IEEEP..97.1482D} radio surveys.

On the other hand optical photometry alone is not fruitful either in separating the populations of associated galaxies because there are degeneracies between the observables of quenching between backsplash orbits and pre-processing of the infalling satellites \citep{2011MNRAS.412..529K}. In the Local Group, availability of proper motion measurements \citep{2021MNRAS.501.2363M,2022A&A...657A..54B} enables us to determine backsplash candidates. Beyond the Local Group, 
 measuring distances using the surface brightness fluctuation (SBF) method has lead to the detection of a backsplash LSBG in \citet{2022arXiv221100629C} with $\mu_{\rm eff,R} = 24.71\,\sbfu$, $r_{\rm eff} = 0.79$\,kpc) that is associated with the M81 galaxy as its host. Named dw0910p7326 this galaxy is apparently composed of a quenched old stellar population with an age of $\lesssim 10$ Gyr. Particularly, LSST is poised to improve SBF based distance measurements with higher quality of data within the first few years of its survey \citep{2021ApJ...908...24G}. Tidal removal of dark matter from the outskirts of backsplash LSBGs can also lead to an increase in their subhalo-stellar mass ratio of these objects. Stacking the dark matter density profiles derived from weak lensing \citep{2018MNRAS.478.1244S} is another way to study the orbital histories of the associated LSBGs in a better fashion.

%More so, the associated LSBGs that we find around the hosts in the Fornax-Eridanus cluster show that even in the high density environment they have their own substructure with pre-processing taking place.
 
%\citep{2021A&A...656A..44R}
%
\subsection{Quenching} \label{sec:quench_disc}

%The fact that we detect the signatures of quenching near the \zm{} hosts across the extremes of environments they are located in favor a mechanism of environmental origin \citep[e.g.][]{2018MNRAS.477.4491F}. 

%The presence of the red sequence demonstrates quenching of the LSBGs in their respective environments. The result that stands out here is that this signature exists even in the low density environment of the associated galaxies around the isolated hosts considered here. Furthermore the galaxies in the green valley that are evolving from the blue cloud to the red sequence is more prominent in the distributions for the isolated environment. This is because these hosts which are out in the field are assembled later compared to the cluster hosts \citep{2007MNRAS.375..489H}. Therefore, alongside fully quenched LSBGs there are those that are star forming and partially processed near the hosts. The population as a whole is not well processed by the isolated hosts in contrast to the cluster hosts. 

The presence of the red-excess in the $\gi{}$ color distributions for the both the satellite and associated LSBGs around cluster and isolated hosts of MW and LMC masses shows the evidence of environmental quenching \citep[e.g.][]{2018MNRAS.477.4491F} processing of the LSBGs. The method of background subtraction and consequent comparison with the distant field sample furthermore rules out the role of internal mechanisms of quenching like stellar feedback or reionization \citep{2017MNRAS.466L...1D,2017MNRAS.465.1682H}. While LSBGs in dense cluster environments are known to be red \citep{2015ApJ...798L..45V,2016A&A...590A..20V,2021A&A...646L..12B,2022arXiv220502193Z} with similar processes of quenching have acted on them as those on higher surface brightness galaxies. On the other hand, LSBGs in the field have been observed to be star-forming \citep{2017ApJ...842..133L} with only $26 \pm 5$ per cent of isolated LSBGs being quiescent \citep{2021MNRAS.500.2049P}.

%LSBGs represent outliers in the mass-size relation (MSR), which is found to be universally obeyed across a range of galaxy masses and independent of of their color and morphology \citep{2021ApJ...922..267C}. If biases imposed by taking a constant cut on the sizes of LSBGs are removed,

To understand the quenching of LSBGs, we also need to know how this is connected to their low surface-brightness nature.  \citet{2023arXiv230214108L} found that the quenched fraction of satellite LSBGs around MW analogs matches that of the broader sample of classical dwarfs \citep{2022ApJ...933...47C}. Therefore, the diffuse nature of LSBGs can be explained as some process that ``puffs up" hitherto normal sized dwarfs \citep{2023arXiv230214108L} causing a vertical movement in the mass-size space at fixed stellar mass. This on top of any reduction in luminosity caused due to quenching, should increase the effective surface brightness (which depends on galaxy size and luminosity). Therefore this process should produce a shift in $\mug$ towards the faint end of the CSB space. Simulations show that this can arise from an internal mechanism like supernovae feedback \citep{2017MNRAS.466L...1D} or environmental processes like tidal heating and ram-pressure stripping \citep{2019MNRAS.487.5272J}. Antlia 2, Crater 2 \citep{2016MNRAS.459.2370T,2019MNRAS.488.2743T,2021ApJ...921...32J} AndXIX, AndXXI, AndXXIII \citep{2016ApJ...833..167M,2020MNRAS.491.3496C,2021MNRAS.505.5686C},  Scl-MM-Dw2 \citep{2022ApJ...926...77M}, and NGC 55-dw1 \citep{2023arXiv230904467M} are examples of recently discovered faint, diffuse satellites whose nature can be attributed to intense tidal stripping by their hosts. Furthermore, \citet{2020MNRAS.494.1848S} and \citet{2022arXiv220907539B} show that among the population of satellite UDGs in the group environment, a fraction were field UDGs before infall while the rest became UDGs after infall aided by tidal heating in the host environment. At the same time there might be multiple mechanisms shaping these objects over the course of their lifetimes \citep{2017A&A...601L..10P}. According to \citet{2021ApJ...920...72K} the shapes of LSBGs are a means to distinguish between the various formation theories that is independent of the LSBG environment. 

%At this point, a pertinent question is are we looking at the same processes when we talk about quenching and puffing? 

The same processes are known to lead to quenching as well. While tidal heating plays a role in quenching, the effect of ram-pressure stripping becomes stronger with a higher ambient density \citep{2019MNRAS.485..796M}. Accordingly, greater ram-pressure stripping is expected to take place closer to the center of the host halo where the density is high. This is the plausible reason for the increase in the quiescent fraction in bright galaxies \citep{2015MNRAS.447L...6S,2022arXiv221003748K} and LSBGs alike \citep{2022arXiv221014237G,2022arXiv221000009K}. The properties of the infalling galaxy also determines the future state of star formation as more massive systems are more resilient to the quenching processes \citep{2022arXiv220813805P,2022arXiv220907539B}. Quenching should correspond to a horizontal movement of galaxies from bluer to redder values of $\gi$ in the CSB space.

The sample of LSBGs inside the Fornax-Eridanus cluster provide the best way to probe these processes in depth. The projected radial distributions show that they are bound to the \zm{} hosts in the same environment. This results in what is known as galactic conformity \citep{2006MNRAS.366....2W} where red dwarfs cluster strongly with respect to the red hosts. The \zm{} hosts that are in the cluster have been inside this environment longer, implying the satellite and associated LSBGs are also old and quenched. However the same LSBGs also show the effects of being part of the larger cluster scale host as well and thereby shaped by the intercluster medium (ICM). This leads to the radial distributions of cluster LSBGs selected by color being relatively flat and the $\gi$ color distributions of the satellite and associated LSBGs in the cluster environment resembling each other closely. We are interested in seeing how galaxies at different mass scales interact amongst themselves. Since galaxies assemble in the mode of hierachical structure formation, it is likely that their quenching takes place along similar lines. This phenomenon of ``hierachical quenching" can be investigated further using low mass cluster substructure \citep{2023arXiv230406886W} like the satellite and associated LSBGs we find in this work.

Beyond the extent of the virial radius of the host galaxy, the same processes that contributes to the quenching of satellites acts to deplete the gas reservoirs of infalling dwarfs. For example, the effect of tidal interactions might extend beyond the virial radius as well \citep{2021MNRAS.506.2766H}. Pre-processing involves a population of dwarfs being subjected to ram-pressure stripping within a low-mass group \citep{2017MNRAS.467.3268R} followed by their collective accretion on to a MW host \citep{2022arXiv221207518S,2022MNRAS.514.5276S}. The LMC can be held responsible for bringing with it its own population of dwarf galaxies that eventually merged with those of the MW \citep{2015MNRAS.453.3568D,2020ApJ...893...48N,2020ApJ...893..121P,2020MNRAS.495.2554E,2021MNRAS.504.5270D,2022MNRAS.513.2673J,2022ApJ...940..136P}. For example, the backsplash LSBG dw0910p7326 interacted with M81 recently ($\sim 1.5$ Gyr ago) compared to the age of its stellar population suggesting that pre-processing played a role in quenching it \citep{2022arXiv221100629C}. Alternative mechanisms of environmental quenching of dwarf galaxies include quenching in cosmic sheets \citep{2022MNRAS.tmp.3505P} and in infall region of galaxy clusters, particularly along filaments \citep{2016MNRAS.455..127M,2022MNRAS.517.4515S}. On the other hand, mergers not only quench galaxy populations but can lead to their diffuse appearances \citep{2021MNRAS.502.5370W,2022MNRAS.511.3210P}.
 
Such processes are better investigated around the isolated hosts because they are the highest mass structure in their vicinity. Unlike the cluster LSBGs, we find differences in the properties of the satellite and associated samples namely sharper gradients in their density profile as well as distinct color distributions. Again in line with galactic conformity, these systems are younger and have been assembled recently \citep{2007MNRAS.375..489H}, therefore the LSBGs contained in them have not been well processed like their counterparts inside the cluster. Around these hosts, it is likely that the red sequence/red-excess we are detecting comprises backsplash LSBGs which have properties similar to the satellite LSBGs while the blue cloud and green valley is made up of the infalling LSBGs subject to pre-processing. Nonetheless, such systems will be ideal to probe those mechanisms of environmental quenching disucssed hitherto that can operate beyond the virial radii of the host galaxies.

%This can also explain the existence of the more extreme population of LSBGs, that are the UDGs. \citet{2022arXiv220802634J} found that there was gradient in $u-i$ colors of LSBGs in the Virgo cluster with bluer ones dominating beyond the virial radius. Also UDGs were found to concentrated mainly towards the center of the cluster.

%A way to address the host-to-host differences, e.g., radial distribution \citep{2020ApJ...902..124C}, in the satellite populations of MW-like hosts is to consider their accretion histories \citep{2021ApJ...909..112D}. 

\section{Conclusion}

In this work we utilize the full catalog of LSBGs from the Y3 DES catalog, seeking to identify their associations with host galaxies having $9.0<{\rm log}(\mathcal{M}_{\star}/M_{\odot})<11.0$ drawn from the low redshift \zm{} sample. The host sample is sub-divided into cluster and isolated environments based on a Friends Of-Friends algorithm with a linking length of 1.5 Mpc. We use the projected virial radius $\theta_{200}$ to define the boundary of the halo containing the host galaxy and select surrounding LSBGs in terms of its projected separations from the hosts. These galaxies are divided into three bins of projected separation from the hosts: $\theta < \theta_{200}$, $ \theta_{200} < \theta < 2.5 \theta_{200}$ and $2.5 \theta_{200} < \theta < 4.5 \theta_{200}$ that we refer to as the satellite, associated and background samples respectively. Our results are as follows:

\begin{itemize}[leftmargin=0.2in]
    \item Computing the background subtracted radial density profile, we find that the LSBGs strongly cluster around the \zm{} host galaxies, and this tendency is enhanced around the hosts with MW-like masses and/or in cluster like environments. Redder and brighter LSBGs are more centrally concentrated around the \zm{} hosts relative to the shallower radial profile of the bluer, brighter LSBGs.
    
    \item We then use a Kernel Density Estimate of the LSBG samples to construct the equivalent Hess diagrams in color-surface brightness space. After background subtraction, there are well-defined red sequences in both the satellite ($\theta < \theta_{200}$) and associated ($ \theta_{200} < \theta < 2.5 \theta_{200}$) regions in both cluster and isolated environments. The latter environment shows pronounced blue cloud and green valley features that suggest ongoing pre-processing.
    
    \item We marginalize these densities along $\mug$ to generated $\gi$ color distributions of the samples and compare with a distant field sample of LSBGs$- $ not located within $2.5\theta_{200}$ of any \zm{} hosts. There is a clear excess of LSBGs at $\gi \sim 0.7$ with respect to the distant field, as substantiated by comparing using a KS-test.
    
    \item These results in combination provide strong support for the existence of quenching in backsplash and infalling LSBGs beyond the classical halo boundary $\theta_{200}$ of the host galaxies. This shows ubiquity of backsplash galaxies across both the environments as well as the importance of pre-processing of infalling galaxies.
\end{itemize}

Our work casts light on the value of low mass LSBGs in tracing how the environment impacts galaxy evolution of the underlying structure. We see similar quenching signatures in both environments which also shows that the centrals with $9.0<{\rm log}(\mathcal{M}_{\star}/M_{\odot})<11.0$ \citep{2022MNRAS.513.2673J} play an important role in quenching/processing the LSBGs around them. 

Investigating systems of either mass scale in the Local Universe beyond the Local Group could herald new discoveries in the field of near-field cosmology. The Legacy Survey of Space and Time (LSST) \citep{2019ApJ...873..111I}, promises to push the current detection limits to even lower surface brightness and greater distance, with a great potential of discovery of LSBG systems in the Local Volume
\citet{2021ApJ...918...88M}. Spectroscopic observations can be useful too, e.g.,
Dragonfly Spectral Line Mapper \citep{2022SPIE12182E..1TL} in gaining velocity information that can be used to discern between backsplash and infalling systems. One of the most promising endeavors will be to study the low mass systems using weak lensing in the Merian Survey \citep{2023arXiv230519310L}.

%While we use color as a the main discriminant for quenching, \citet{2022MNRAS.510.4126S} show that galaxy color does not perfectly match with morphology which is directly determined by the star formation history. 

\section*{Acknowledgment}

This paper has gone through internal review by the DES collaboration. The authors would like to thank Adam Leroy, Jenny E. Greene and Benedikt Diemer for their valuable input. JB and AHGP are supported by National Science Foundation Grant No. AST-2008110.

Funding for the DES Projects has been provided by the U.S. Department of Energy, the U.S. National Science Foundation, the Ministry of Science and Education of Spain, the Science and Technology Facilities Council of the United Kingdom, the Higher Education Funding Council for England, the National Center for Supercomputing Applications at the University of Illinois at Urbana-Champaign, the Kavli Institute of Cosmological Physics at the University of Chicago, the Center for Cosmology and Astro-Particle Physics at the Ohio State University, the Mitchell Institute for Fundamental Physics and Astronomy at Texas A\&M University, Financiadora de Estudos e Projetos, Funda{\c c}{\~a}o Carlos Chagas Filho de Amparo {\`a} Pesquisa do Estado do Rio de Janeiro, Conselho Nacional de Desenvolvimento Cient{\'i}fico e Tecnol{\'o}gico and the Minist{\'e}rio da Ci{\^e}ncia, Tecnologia e Inova{\c c}{\~a}o, the Deutsche Forschungsgemeinschaft and the Collaborating Institutions in the Dark Energy Survey. The Collaborating Institutions are Argonne National Laboratory, the University of California at Santa Cruz, the University of Cambridge, Centro de Investigaciones Energ{\'e}ticas, Medioambientales y Tecnol{\'o}gicas-Madrid, the University of Chicago, University College London, the DES-Brazil Consortium, the University of Edinburgh, the Eidgen{\"o}ssische Technische Hochschule (ETH) Z{\"u}rich, Fermi National Accelerator Laboratory, the University of Illinois at Urbana-Champaign, the Institut de Ci{\`e}ncies de l'Espai (IEEC/CSIC), the Institut de F{\'i}sica d'Altes Energies, Lawrence Berkeley National Laboratory, the Ludwig-Maximilians Universit{\"a}t M{\"u}nchen and the associated Excellence Cluster Universe, the University of Michigan, NSF's NOIRLab, the University of Nottingham, The Ohio State University, the University of Pennsylvania, the University of Portsmouth, SLAC National Accelerator Laboratory, Stanford University, the University of Sussex, Texas A\&M University, and the OzDES Membership Consortium. Based in part on observations at Cerro Tololo Inter-American Observatory at NSF's NOIRLab (NOIRLab Prop. ID 2012B-0001; PI: J. Frieman), which is managed by the Association of Universities for Research in Astronomy (AURA) under a cooperative agreement with the National Science Foundation. The DES data management system is supported by the National Science Foundation under Grant Numbers AST-1138766 and AST-1536171. The DES participants from Spanish institutions are partially supported by MICINN under grants ESP2017-89838, PGC2018-094773, PGC2018-102021, SEV-2016-0588, SEV-2016-0597, and MDM-2015-0509, some of which include ERDF funds from the European Union. IFAE is partially funded by the CERCA program of the Generalitat de Catalunya. Research leading to these results has received funding from the European Research Council under the European Union's Seventh Framework Program (FP7/2007-2013) including ERC grant agreements 240672, 291329, and 306478. We  acknowledge support from the Brazilian Instituto Nacional de Ci\^encia e Tecnologia (INCT) do e-Universo (CNPq grant 465376/2014-2). This manuscript has been authored by Fermi Research Alliance, LLC under Contract No. DE-AC02-07CH11359 with the U.S. Department of Energy, Office of Science, Office of High Energy Physics.

\textit{Software:} \texttt{NumPy} \citep{2020Natur.585..357H}, \texttt{Matplotlib} \citep{2007CSE.....9...90H}, \texttt{SciPy} \citep{2020SciPy-NMeth}, \texttt{astropy} \citep{2013A&A...558A..33A,2018AJ....156..123A}, \texttt{Source Extractor} \citep{1996A&AS..117..393B}.

\section*{Author Affiliations}
$^{1}$ Department of Astronomy, The Ohio State University, Columbus, OH 43210, USA\\
$^{2}$ Center for Cosmology and Astro-Particle Physics, The Ohio State University, Columbus, OH 43210, USA\\
$^{3}$ Department of Physics, The Ohio State University, Columbus, OH 43210, USA\\
$^{4}$ Institute for Advanced Study, 1 Einstein Drive, Princeton, NJ 08540, USA\\
$^{5}$ Dartmouth College, Department of Physics and Astronomy, Hanover, NH 03755, USA\\
$^{6}$ Department of Astronomy and Astrophysics, University of Chicago, Chicago, IL 60637, USA\\
$^{7}$ Kavli Institute for Cosmological Physics, University of Chicago, Chicago, IL 60637, USA\\
$^{8}$ Fermi National Accelerator Laboratory, P.O. Box 500, Batavia, IL 60510, USA\\
$^{9}$ McWilliams Center for Cosmology, Carnegie Mellon University, 5000 Forbes Ave, Pittsburgh, PA 15213, USA\\
$^{10}$ Texas A\&M University, College Station, TX 77843, USA\\
$^{11}$ Centre for Extragalactic Astronomy, Durham University, South Road, Durham DH1 3LE, UK\\
$^{12}$ Laborat\'orio Interinstitucional de e-Astronomia - LIneA, Rua Gal. Jos\'e Cristino 77, Rio de Janeiro, RJ - 20921-400, Brazil\\
$^{13}$ Department of Physics, University of Michigan, Ann Arbor, MI 48109, USA\\
$^{14}$ Institute of Cosmology and Gravitation, University of Portsmouth, Portsmouth, PO1 3FX, UK\\
$^{15}$ Department of Physics \& Astronomy, University College London, Gower Street, London, WC1E 6BT, UK\\
$^{16}$ Instituto de Astrofisica de Canarias, E-38205 La Laguna, Tenerife, Spain\\
$^{17}$ Universidad de La Laguna, Dpto. Astrofísica, E-38206 La Laguna, Tenerife, Spain\\
$^{18}$ Institut de F\'{\i}sica d'Altes Energies (IFAE), The Barcelona Institute of Science and Technology, Campus UAB, 08193 Bellaterra (Barcelona) Spain\\
$^{19}$ Hamburger Sternwarte, Universit\"{a}t Hamburg, Gojenbergsweg 112, 21029 Hamburg, Germany\\
$^{20}$ School of Mathematics and Physics, University of Queensland,  Brisbane, QLD 4072, Australia\\
$^{21}$ Department of Physics, IIT Hyderabad, Kandi, Telangana 502285, India\\
$^{22}$ Institute of Theoretical Astrophysics, University of Oslo. P.O. Box 1029 Blindern, NO-0315 Oslo, Norway\\
$^{23}$ Fermi National Accelerator Laboratory, P. O. Box 500, Batavia, IL 60510, USA\\
$^{24}$ Instituto de Fisica Teorica UAM/CSIC, Universidad Autonoma de Madrid, 28049 Madrid, Spain\\
$^{25}$ University Observatory, Faculty of Physics, Ludwig-Maximilians-Universit\"at, Scheinerstr. 1, 81679 Munich, Germany\\
$^{26}$ Center for Astrophysical Surveys, National Center for Supercomputing Applications, 1205 West Clark St., Urbana, IL 61801, USA\\
$^{27}$ Department of Astronomy, University of Illinois at Urbana-Champaign, 1002 W. Green Street, Urbana, IL 61801, USA\\
$^{28}$ Santa Cruz Institute for Particle Physics, Santa Cruz, CA 95064, USA\\
$^{29}$ Center for Astrophysics $\vert$ Harvard \& Smithsonian, 60 Garden Street, Cambridge, MA 02138, USA\\
$^{30}$ Australian Astronomical Optics, Macquarie University, North Ryde, NSW 2113, Australia\\
$^{31}$ Lowell Observatory, 1400 Mars Hill Rd, Flagstaff, AZ 86001, USA\\
$^{32}$ George P. and Cynthia Woods Mitchell Institute for Fundamental Physics and Astronomy, and Department of Physics and Astronomy, Texas A\&M University, College Station, TX 77843,  USA\\
$^{33}$ LPSC Grenoble - 53, Avenue des Martyrs 38026 Grenoble, France\\
$^{34}$ Instituci\'o Catalana de Recerca i Estudis Avan\c{c}ats, E-08010 Barcelona, Spain\\
$^{35}$ Department of Physics, Carnegie Mellon University, Pittsburgh, Pennsylvania 15312, USA\\
$^{36}$ Observat\'orio Nacional, Rua Gal. Jos\'e Cristino 77, Rio de Janeiro, RJ - 20921-400, Brazil\\
$^{37}$ Kavli Institute for Particle Astrophysics \& Cosmology, P. O. Box 2450, Stanford University, Stanford, CA 94305, USA\\
$^{38}$ SLAC National Accelerator Laboratory, Menlo Park, CA 94025, USA\\
$^{39}$ Centro de Investigaciones Energ\'eticas, Medioambientales y Tecnol\'ogicas (CIEMAT), Madrid, Spain\\
$^{40}$ Instituto de F\'\i sica, UFRGS, Caixa Postal 15051, Porto Alegre, RS - 91501-970, Brazil\\
$^{41}$ School of Physics and Astronomy, University of Southampton,  Southampton, SO17 1BJ, UK\\
$^{42}$ Computer Science and Mathematics Division, Oak Ridge National Laboratory, Oak Ridge, TN 37831\\
$^{43}$ Cerro Tololo Inter-American Observatory, NSF's National Optical-Infrared Astronomy Research Laboratory, Casilla 603, La Serena, Chile\\
$^{44}$ Lawrence Berkeley National Laboratory, 1 Cyclotron Road, Berkeley, CA 94720, USA

\bibliography{sample631}{}
\bibliographystyle{aasjournal}

\end{document}